\documentclass[twocolumn,showpacs,preprintnumbers,amsmath,amssymb,floatfix]{revtex4-1}
\usepackage{mysty}

\begin{document}

\title{Direct numerical simulation of single mode three-dimensional Rayleigh-Taylor experiments}

\author{Maxwell Hutchinson}
\affiliation{The Physics Department, University of Chicago, Chicago IL 60637}
\email{maxhutch@uchicago.edu}

\date{\today}

\begin{abstract}
The single-mode Rayleigh-Taylor instability (smRTI) is well defined, poorly understood, and applicable to many fluid flows directly and through its relationship to multi-mode Rayleigh-Taylor models.
This study reproduces three low-Atwood smRTI experimental runs (Wilkinson and Jacobs, 2007) in
a specialized version of the Nek5000 spectral element code.
The simulations use the initial amplitude, wavelength, acceleration, Atwood number, and viscosity from 
the three specific experiments and impose no-slip and no-flux boundaries on the velocity and scalar, respectively.
The simulations are shown to reproduce the linear, saturation, stagnation, and re-acceleration phases of the smRTI
seen in the experiments.
Additionally, access to the full velocity and scalar fields demonstrates three different finite size effects: 
wall drag, wall lift, and a long wavelength mode along the diagonal.
One of the simulations is extended by a factor of two in the vertical direction and the resulting late-time dynamics
reach Froude numbers around 1.8, higher than previously reported.
Finally, inspection of the span-wise flow reveals secondary flows of the first kind that transport the scalar 
from the bubble-spike interfaces into the bubble and spike centers.
The agreement between simulations and experiments inspires confidence in the spectral element method for studying
the Rayleigh-Taylor instability.
\end{abstract}

\maketitle

\makeatletter{}\makeatletter{}\section{Introduction} \slabel{intro}

The Rayleigh-Taylor instability occurs when a denser fluid is supported by a lighter one.
Low-amplitude perturbations in the interface between the two fluids grow exponentially with a rate that is well modeled by linear stability analysis~\cite{Duff1962}:
\begin{equation} \elabel{duff}
\gamma = \sqrt{\frac{Agk}{\psi} + \nu^2 k^4} - (\nu + D) k^2
\end{equation}
where the Atwood number $A = \Delta \rho / \sum \rho$,
$g$ is the local acceleration,
$k$ is the wave-number,
$\nu$ is the kinematic viscosity,
$D$ is the diffusivity,
and $\psi \approx 1$ describes the effect of the interface thickness and is a function of $A$, $k$, and the thickness $\delta$.
In the low Atwood number limit, $\psi = 1 + \pi^{-1/2} k \delta$.

At larger amplitudes, the perturbations grow non-linearly with the light fluid rising through the heavier fluid in `bubbles' and the heavy fluid falling through the lighter fluid in `spikes.'
Early experiments by Davies and Taylor~\cite{Davies1950a} and potential flow models by Layzer~\cite{Layzer1955} for $A \approx 1$ suggested that the bubbles reach a terminal velocity, and later experiments by Dimonte and Schneider~\cite{Dimonte1996}, also at $A \approx 1$, showed that dense spikes free-fall.
On the other hand, recent experiments by Wilkinson and Jacobs~\cite{Wilkinson2007} and simulations by Ramaprabhu et al.~\cite{Ramaprabhu2006,Ramaprabhu2012}, Wei and Livescu~\cite{Wei2012}, and others~\cite{Sohn2011} show that at smaller density contrasts, the constant velocity regime is followed by a re-acceleration regime in which the velocity doubles.
The dynamics beyond re-acceleration have not been established, with Ramaprabhu et al. observing a return to the velocity of potential flow~\cite{Goncharov2002} while Wei and Livescu report continued constant acceleration.

Here, we consider the low-Atwood number limit, where we can apply the Boussinesq approximation to simplify the governing equations:
\begin{align} \elabel{boussinesq}
\frac{D}{D t} u &= - \nabla P + \nu \nabla^2 u - A \vec{g} \phi, \\
\frac{D}{Dt} \phi &= D \nabla^2 \phi, \nonumber
\end{align}
where $u$ is the velocity,
$P$ is the pressure, and 
$\phi$ is the scalar that controls the density gradient.
These equations have a symmetry under inversion of the scalar and acceleration, $\phi \rightarrow -\phi, \hat{g} \rightarrow -\hat{g}$, so we know the bubbles and spikes have the same dynamics given the same initial conditions.
The parameter space of the equations are described by two non-dimensional numbers: the Grashof number,
\begin{equation} \elabel{grashof}
\text{Gr} = \frac{A g \lambda^3}{\nu^2},
\end{equation}
where $\lambda$ is a characteristic length, and
the Schmidt number,
\begin{equation} \elabel{schmidt}
\text{Sc} = \frac{\nu}{D}
\end{equation}

We approximate the governing equations numerically using the spectral element method (SEM)~\cite{Deville2002}.
The SEM converges exponentially with respect to spectral order and has purely dispersive errors, making it a natural method for direct numerical simulations of mixing problems.
Unlike pseudo-spectral methods, it handles no-slip boundaries and can evenly sample the interior of the domain.
We use a specialized version of the Nek5000 community code, NekBox~\cite{Hutchinsonb}, which restricts the domain to a tensor product of orthogonal bases and employs fast spectral coarse preconditioners.

It is the goal of this study to validate direct numerical simulations of the low-Atwood single mode Rayleigh-Taylor instability in NekBox against the best available experimental data, that from Wilkinson and Jacobs~\cite{Wilkinson2007}.
In the future, we will apply the same numerical methods to the broader question of the late-time dynamics of the low Atwood smRTI.

\paragraph{Outline}
In \sref{methods}, we review the spectral element method and describe the numerical parameters of the simulations.
In \sref{results}, we compare the numerical and experimental results, extend the domain in extent and time to reach higher aspect ratios, and introduce new behavior in the span-wise flow.
In \sref{concs}, we discuss the validity of our methods for simulating Rayleigh-Taylor flows, the limits of wall-bounded single mode experiments, and the implications of secondary flows to mixing.

\makeatletter{}\section{Numerical methods} \slabel{methods}

\subsection{Spectral element formulation}
The governing equations, \eref{boussinesq}, are discritized using the spectral element method~\cite{Deville2002}.
The domain is first divided into cubic elements.
Each element is represented by a tensor product of $n^3$ Gauss-Lobatto-Legendre (GLL) quadtrature points of order $p = n-1$.
The elements are coupled by continuity at the shared points on their faces.

Time is discritized with a 3rd order backwards difference formula (BDF3).
The linear and non-linear terms are split, and the non-linear convection operator is extrapolated with a 3rd order scheme that sets the 3rd derivative at the extrapolated point to zero (EX3).
The full discrete system is 3rd order in time, $p$th order in space, and has purely dispersive errors.

Initially, no filtering is applied.
Filtering is added when the scalar field, which is less stable for Schmidt numbers greater than unity, exhibits small scale oscillations on what should otherwise be smooth steep boundaries.
The filter attenuates the highest frequency mode within each element by 5\% at each time-step.

\subsection{Post-processing}
The simulation outputs the velocity, pressure, and scalar fields at the GLL points in double precision.
They are post-processed into low-dimensional observables: the bubble height and two-dimensional slices of the velocity, vorticity, pressure, and scalar through the horizontal mid-plane and vertical diagonal.
 
The bubble height is defined as:
\begin{equation} \elabel{h_exp}
H = \sup \left\{ z : \min_{x,y} \phi(x,y,z) < 0\right\}
\end{equation}
The bubble velocity is found by fitting a cubic spline to $H(t)$ and differentiating.

The height of individual bubbles, $H_{i,j}$, is found by restricting $\min_{x,y}$ in \eref{h_exp} to the span-wise square of diagonal length $\lambda$ centered on the bubble in the $i$-by-$j$-th position.
The bubble domains and labels are shown in \fref{ic}.

Two-dimensional slices of the velocity, pressure, and scalar are taken in the horizontal mid-plane and in the vertical plane across the diagonal.
The vertical component of the vorticity is computed in the horizontal mid-plane, with derivatives evaluated with spectral accuracy.

Post-processing is performed using the `nek-analyze' post-processing framework, which implements a MapReduce-like backend for parallel, out-of-core analysis.

\subsection{Simulation setup}

\begin{figure}
\includegraphics[width=\columnwidth]{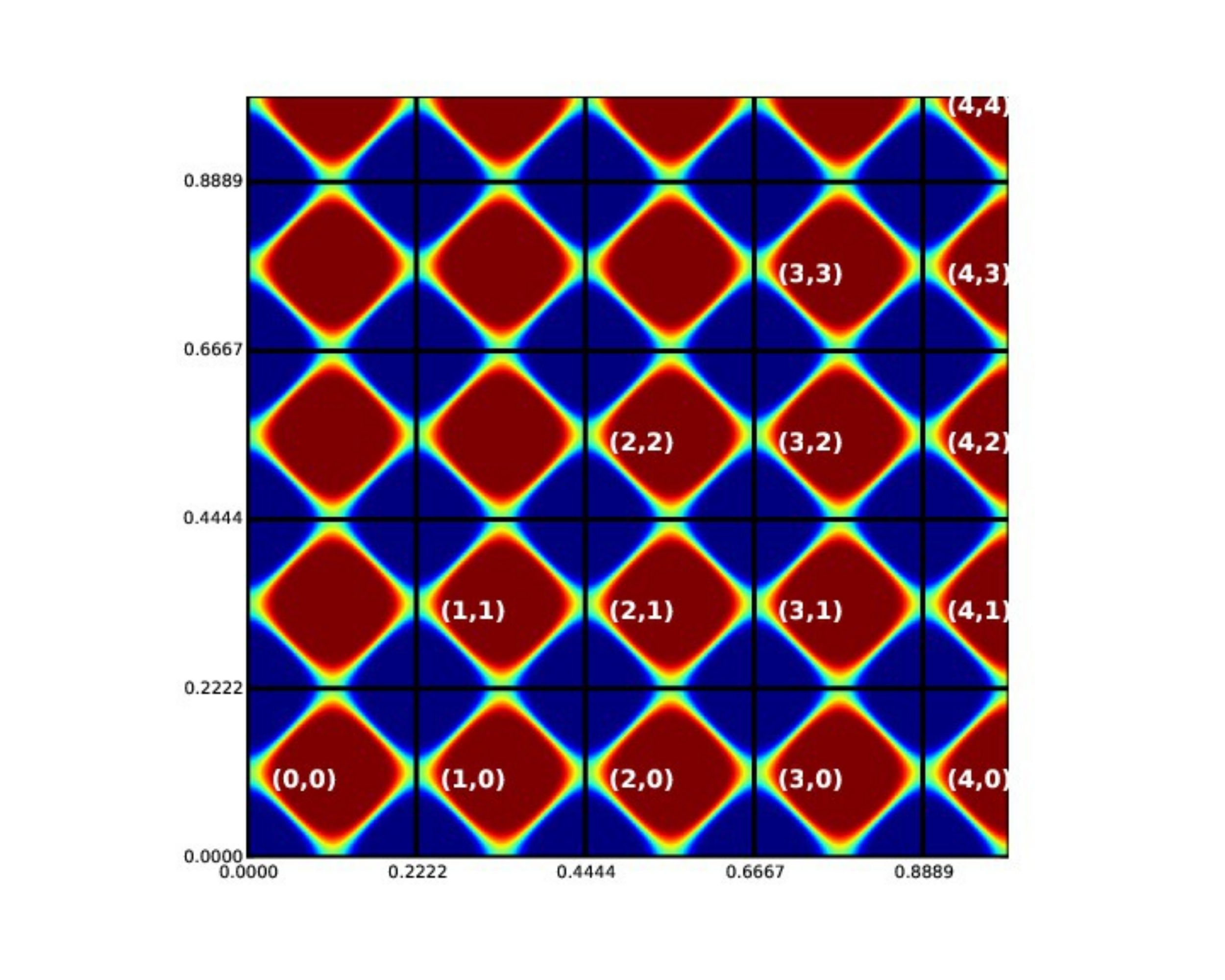}
\caption{ \flabel{ic} 
Initial condition, $\phi(x,y,0,0)$, for 4.5 mode simulation.  
The bubbles are logically separated by the solid grid and each unique bubble is labeled.
}
\end{figure}

\begin{table}
\makeatletter{}\begin{tabular}{rrrr}
\toprule
 Modes &           Grashof &  Schmidt &  Aspect \\
\midrule
   2.5 & $5.12 \times 10^{6}$ &      7.0 &   0.037 \\
   3.5 & $3.98 \times 10^{6}$ &      3.5 &   0.025 \\
   4.5 & $4.66 \times 10^{5}$ &      1.0 &   0.041 \\
\hline
   4.5 & $4.66 \times 10^{5}$ &      3.5 &   0.041 \\
\bottomrule
\end{tabular}
 
\caption{ \tlabel{params}
Parameters of simulations.
The aspect ratio is defined with repsect to the quiecent amplitude, $a_0$, from \eref{rollback}.
The last row is the periodic reference calculation.
}
\end{table}

Five simulations were conducted: reproductions of 2.5, 3.5, and 4.5 mode experiments, an extension of the 4.5 mode experiment with twice the vertical extent, and reference 4.5 mode calculation with periodic boundary conditions.
The boundary conditions for the non-periodic simulations are no-slip for the velocity and no-flux (insulating) for the scalar.
The initial conditions for the velocity are quiescent.
The initial conditions for the scalar are the product of two cosines smeared by an error function in the z-direction:
\begin{equation}
  \phi(x,y,z,0) = \text{erf}\left[ \frac{z + a_0 \cos(k_x x) \cos(k_y y)}{\delta} \right] ,
\end{equation}
where $a_0$ is the initial amplitude,
$k_x$ and $k_y$ are the wave-numbers in the x and y directions, respectively, and
$\delta$ is the interface thickness.
In our case, $k_x = k_y$.
An example initial condition is plotted in \fref{ic}.
By symmetry, we know $H_{i,j} = H_{j,i}$ and that the spike dynamics are identical to the bubbles.

The Atwood number, local acceleration, and initial amplitude where taken from experimental measurements by Wilkinson and Jacobs~\cite{Wilkinson2007}.
In the experiment, the interface has a non-zero initial velocity, so the linear theory, \eref{duff}, is used to transform the flow to quiescence.
Specifically, the system
\begin{align} \elabel{rollback}
H &= a_0 \cosh\left(\gamma (t - t_0)\right) \\
V &= a_0 \gamma \cosh\left(\gamma (t-t_0) \right)  \nonumber
\end{align}
where $H$ and $V$ were the experimental measures of the initial interface height and velocity, respectively,
is solved for $a_0$ and $t_0$, the quiescent amplitude and time.
The quiescent amplitude is used for the simulations.

The three simulations reproducing experimental runs are conducted on the Mira supercomputer at Argonne Leadership Computing Facility (ALCF) using 33,554,432 7th order elements for 11,509,170,176 degrees of freedom distributed over 524,288 cores and 1,048,576 MPI processes.
64 outputs are written to disk, each 6/8ths of a TiB.
The number of elements and degrees of freedom are doubled for the extension to twice the vertical extent.

\makeatletter{}\section{Results} \slabel{results}

\subsection{Validation}

\subsubsection{Linear growth rate}

\begin{table}
\makeatletter{}\begin{tabular}{rrrr}
\toprule
 Modes &  Theory &  Simulation &  Aspect ratio \\
\midrule
   2.5 &   12.94 &       12.96 &         0.089 \\
   3.5 &   22.24 &       22.24 &         0.098 \\
   4.5 &   12.10 &       11.05 &         0.124 \\
\hline
   4.5 &   12.31 &       11.52 &         0.128 \\
\bottomrule
\end{tabular}
 
\caption{ \tlabel{linear}
Growth rate: linear theory vs simulation.
Theoretical values are calculated as in \eref{duff}.
Simulation values are calculated as in \eref{linear_sim}.
The aspect ratio is shown for the second sample, $h_1 / \lambda$.
Note the difference in Schmidt number between the two $4.5$ mode cases.
}
\end{table}

We compute the growth rate from the bubble height at the first simulation output time:
\begin{equation} \elabel{linear_sim}
\gamma \approx \frac{1}{t_1} \text{acosh}\left(\frac{h(t_1)}{h(0)}\right), 
\end{equation}
and collect the results in \tref{linear}.
Because the simulations targeted the non-linear regime, the height is not available until characteristic time $\tau = t \gamma \sim 2$ and aspect ratio $h / \lambda \approx 0.1$.
Therefore, we expect the simulation value to be below the theoretical value given by \eref{duff} due to saturation.

The agreement is good, with only the lower Grashof, higher aspect ratio 4.5 mode calculation deviating more than a part in one hundred.
The 2.5 mode simulation outperforms the theory slightly.
This could be due to the long-wavelength finite size effect discussed later, which is stronger for fewer modes in the finite domain.

\subsubsection{Froude number}

\begin{figure}
\includegraphics[width=\columnwidth]{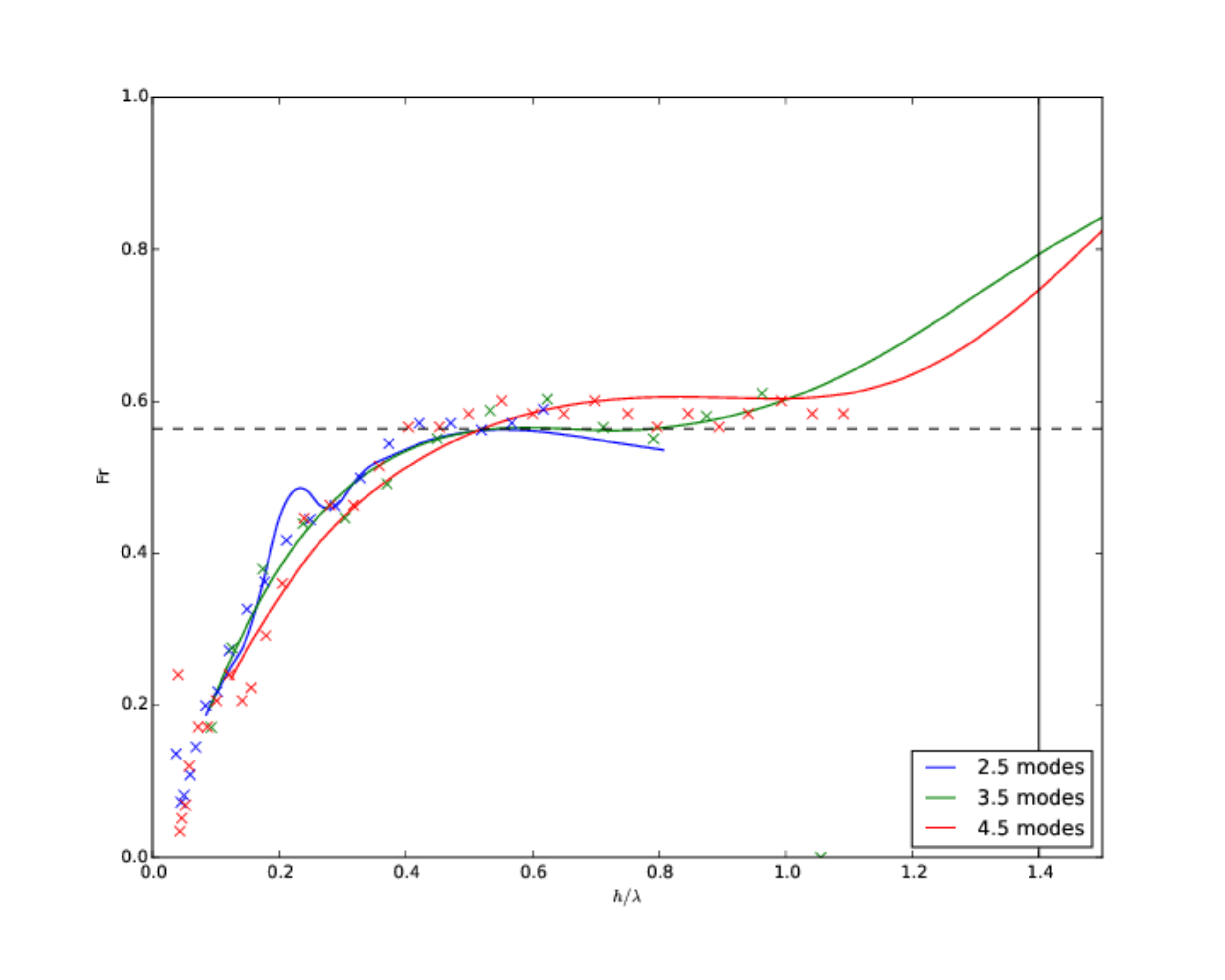}
\caption{ \flabel{froude} 
Froude number vs height, non-dimensionalized by the wavelength.
Lines are the derivative of cubic splines through simulation outputs.
Points are from experiment via direct measurement of the bubble velocity.
The dotted horizontal line is positioned at Goncharov's theoretical value of $\pi^{-1/2}$~\cite{Goncharov2002}.
}
\end{figure}

\begin{figure}
\includegraphics[width=\columnwidth]{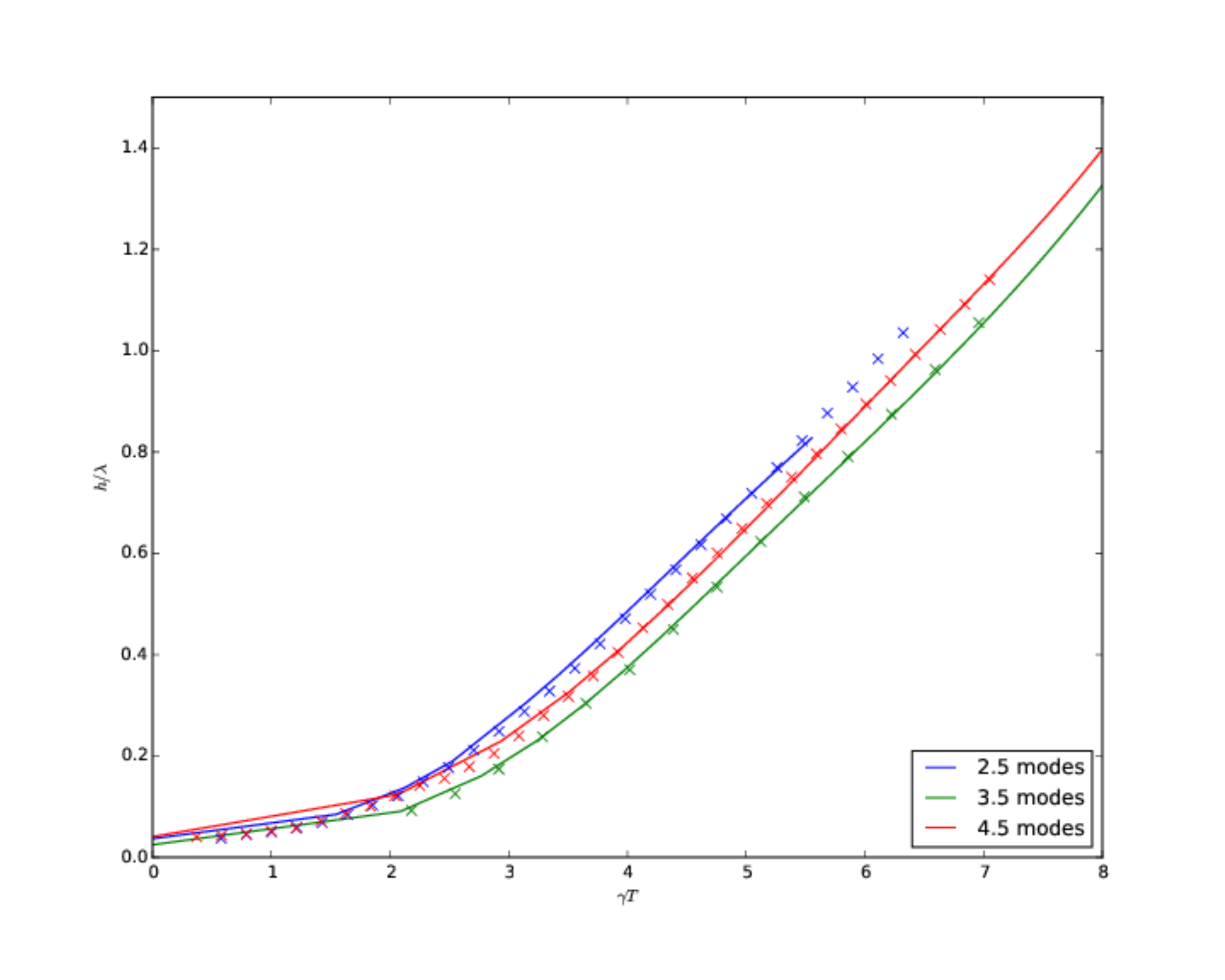}
\caption{ \flabel{aspect} 
Bubble height vs time, non-dimensionalized by the wavelength and linear growth rate.
Lines are cubic splines through simulation outputs.
Points are from experiment.
The points are shifted in time to minimize the square deviation from the spline summed over the plotted points.
}
\end{figure}

The experiments observe only the diagonal plane and measure the height with respect to the most internal bubble.
Therefore, we plot in \fref{froude} the Froude number of the central bubble alone.
In each case, the simulation exhibits the same qualitative behavior as the experiment:
exponential growth saturating to a stagnation velocity around Goncharov's theoretical value of $\pi^{-1/2}$~\cite{Goncharov2002}.
In cases where the simulation and experimental data extend in time, the beginnings of re-acceleration are also seen.

\subsection{Extension}

We extend the study of Wilkinson and Jacobs in two ways.
First, we calculate the height of the full set of bubbles and spikes.
The bubble trajectories are compared to a periodic flow with the same material parameters and initial conditions to characterize the effect of the walls.

Second, we extend the 4.5 mode case by a factor of two in the vertical extent of the domain and simulation time.
The longer trajectories demonstrate late time, high aspect ratio behavior of a flow that would be difficult to realize experimentally.

\subsubsection{Wall effects}

\begin{figure*}
\subfloat[Full trajectory]{
  \includegraphics[width=\columnwidth]{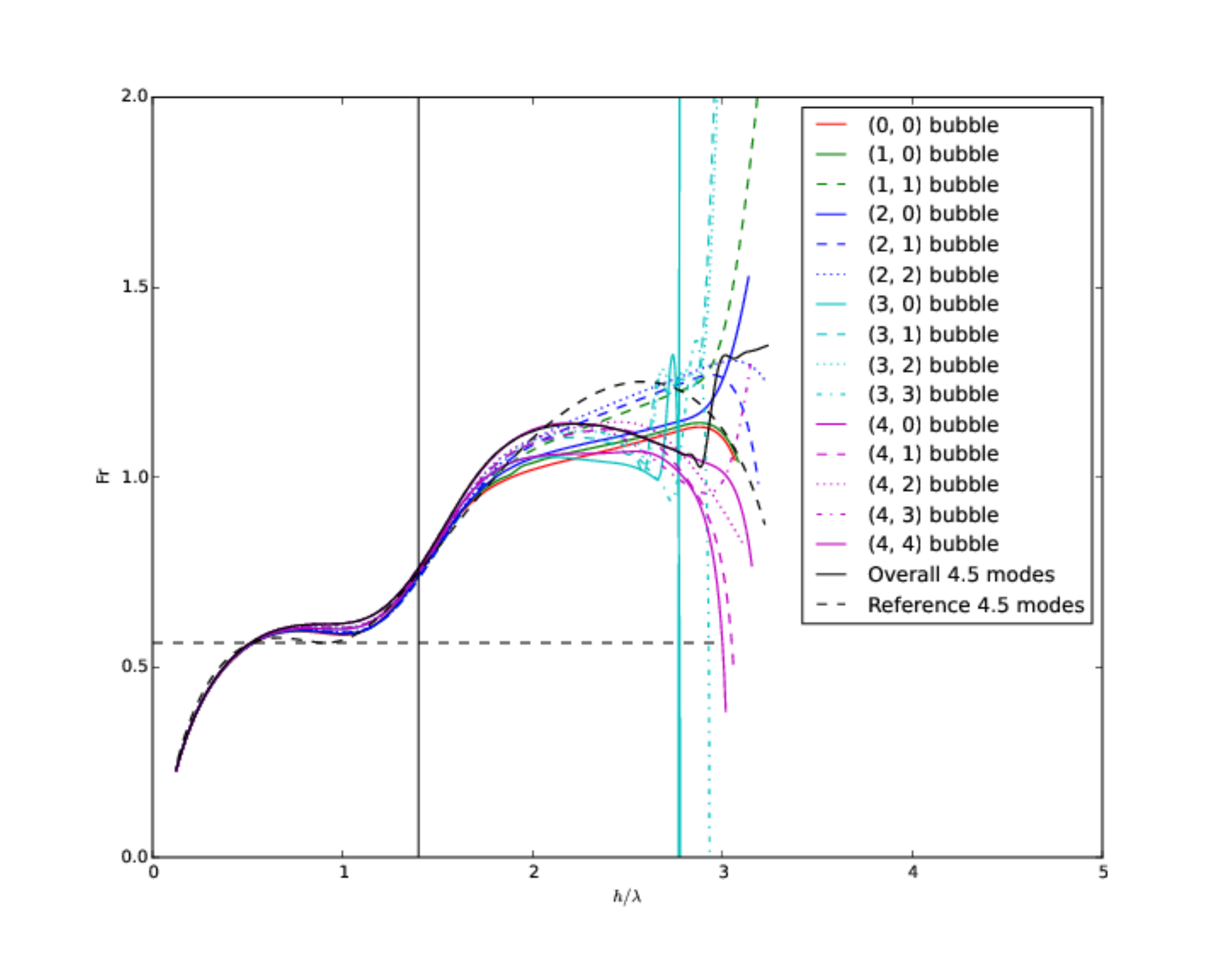}
}
\subfloat[Stagnation regime]{
  \includegraphics[width=\columnwidth]{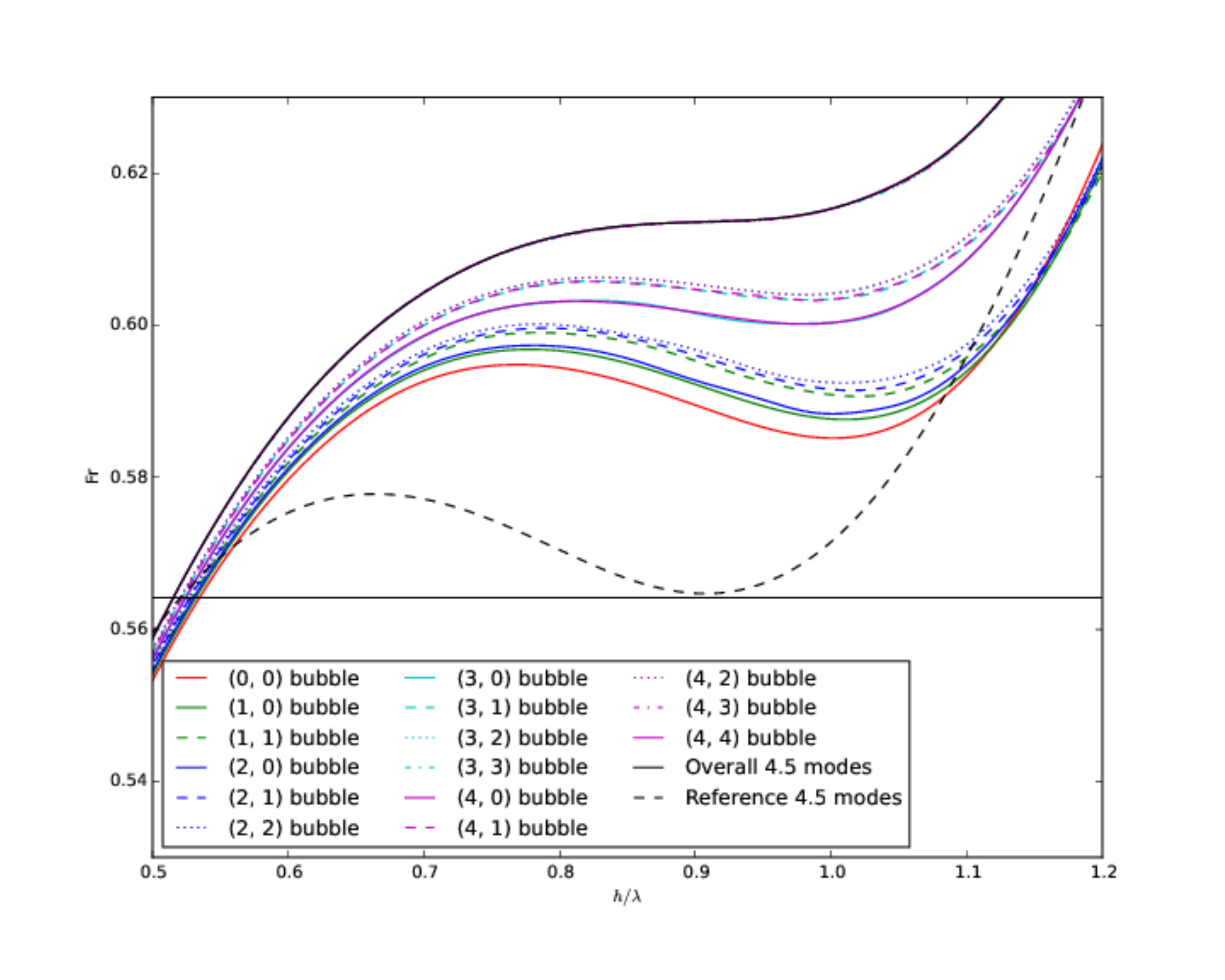}
}
\caption{ \flabel{froude_wall} 
Froude number as a function of height, non-dimensionalized by the wavelength, by bubble in the 4.5 mode simulation.
Solid line is from the height defined as the maximum taken over the entire span-wise domain.
Dotted line is the periodic reference calculation.
}
\end{figure*}

\begin{figure*}
\subfloat[All bubbles]{
  \includegraphics[width=\columnwidth]{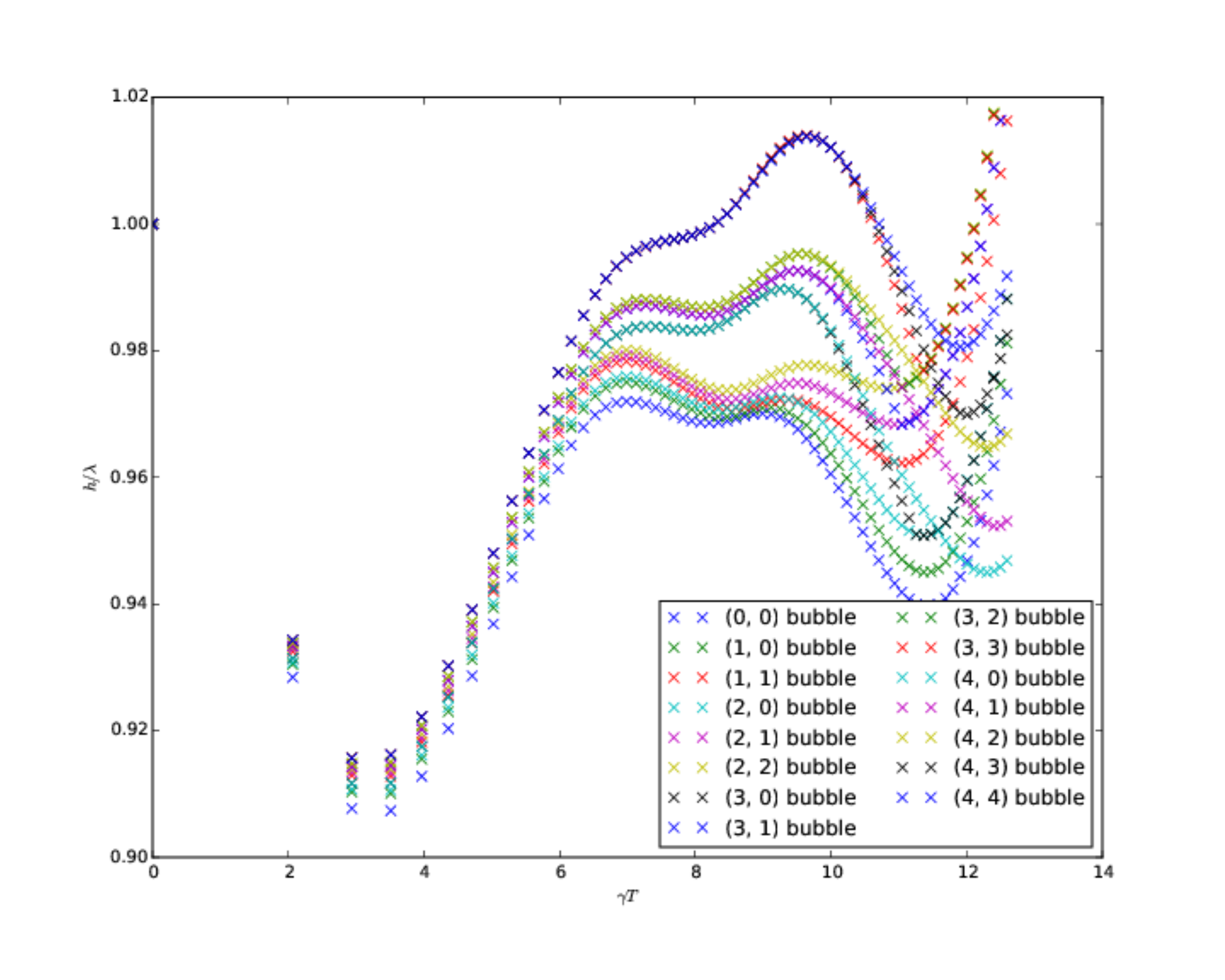}
}
\subfloat[Diagonal bubbles]{
  \includegraphics[width=\columnwidth]{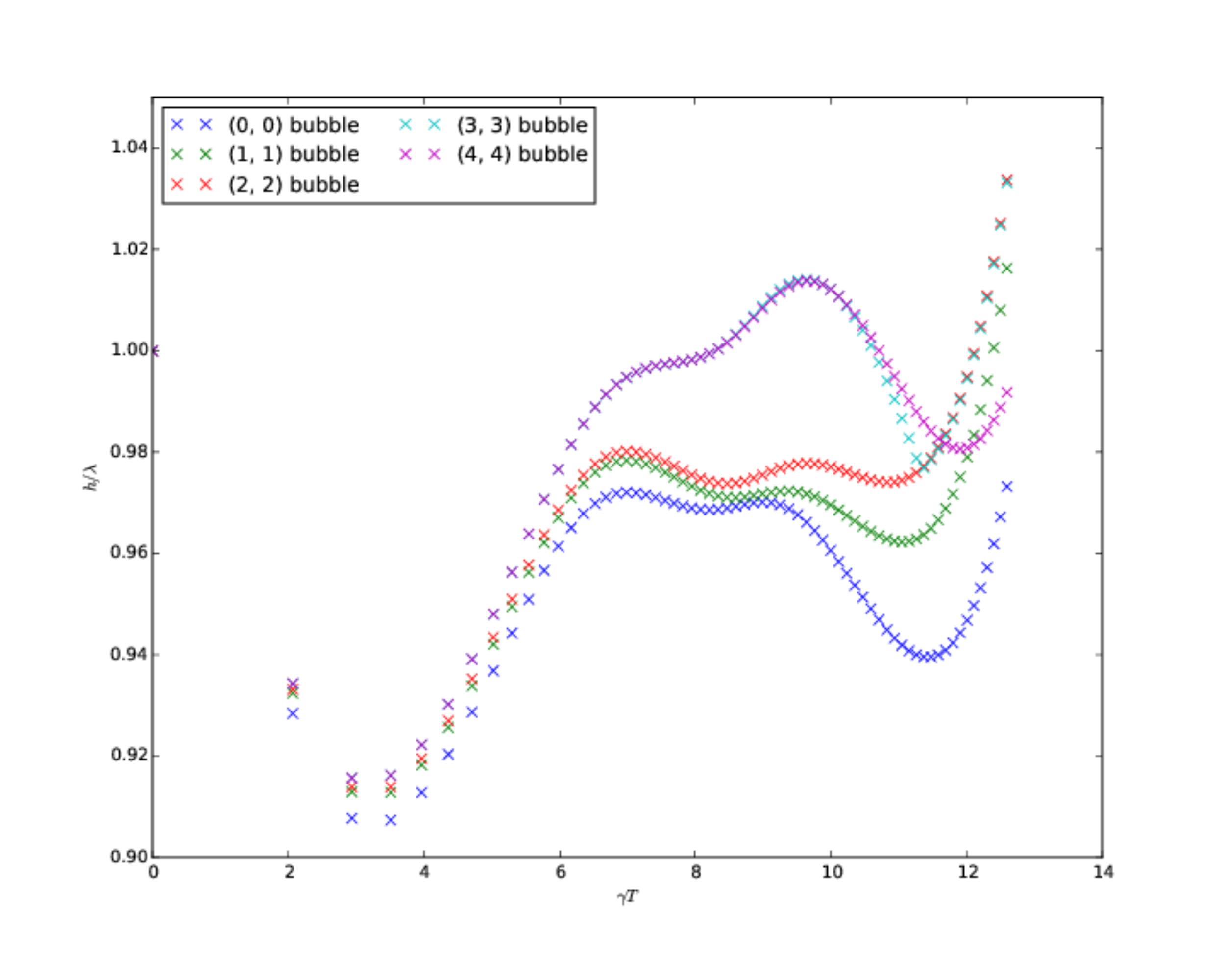}
}
\caption{ \flabel{ratio_wall}
Ratio of wall-bounded bubble height to periodic bubble height in the 4.5 mode simulation.
}
\end{figure*}

\begin{figure*}
\subfloat[Linear regime]{
  \includegraphics[width=0.66\columnwidth]{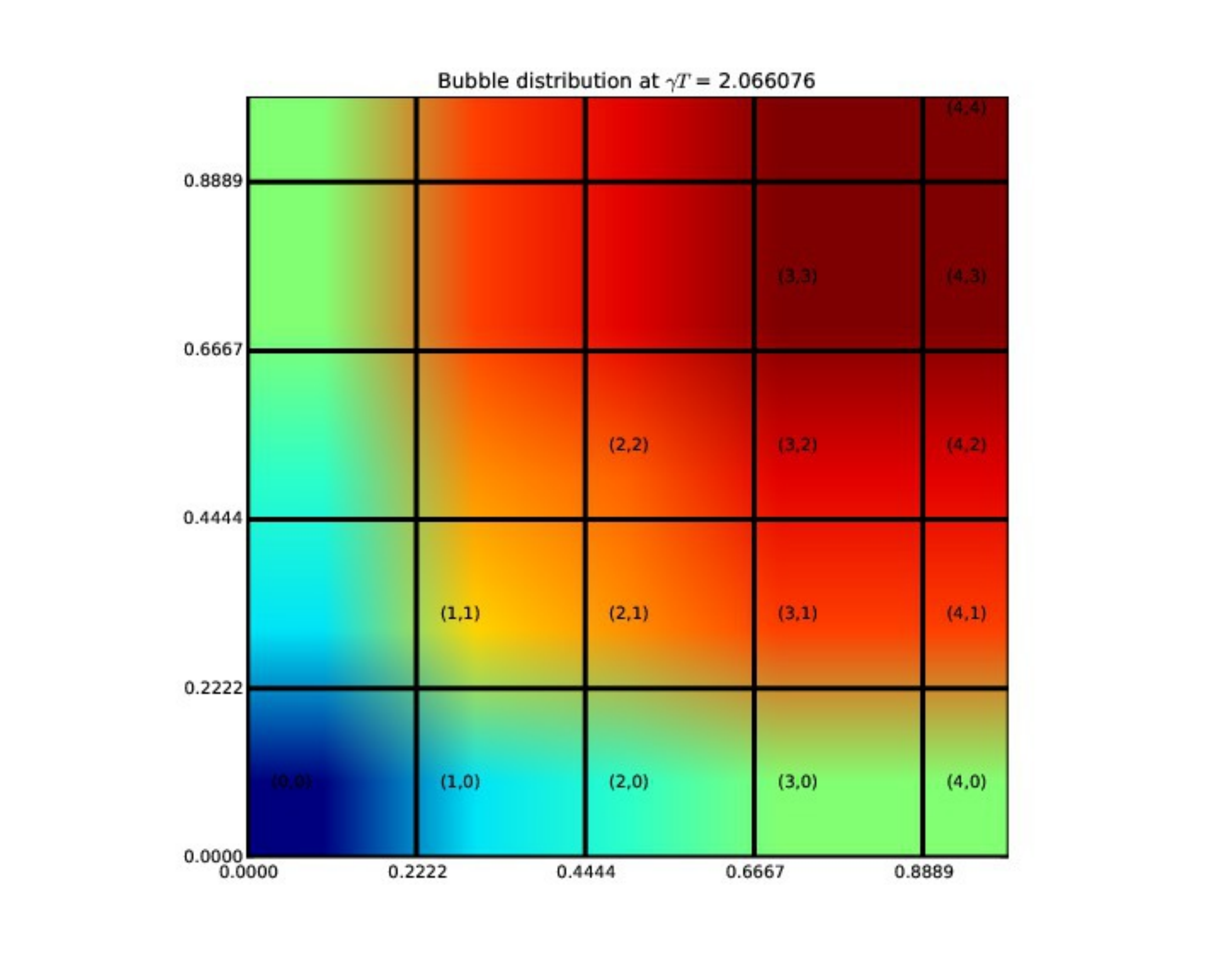}
}
\subfloat[Stagnation regime]{
  \includegraphics[width=0.66\columnwidth]{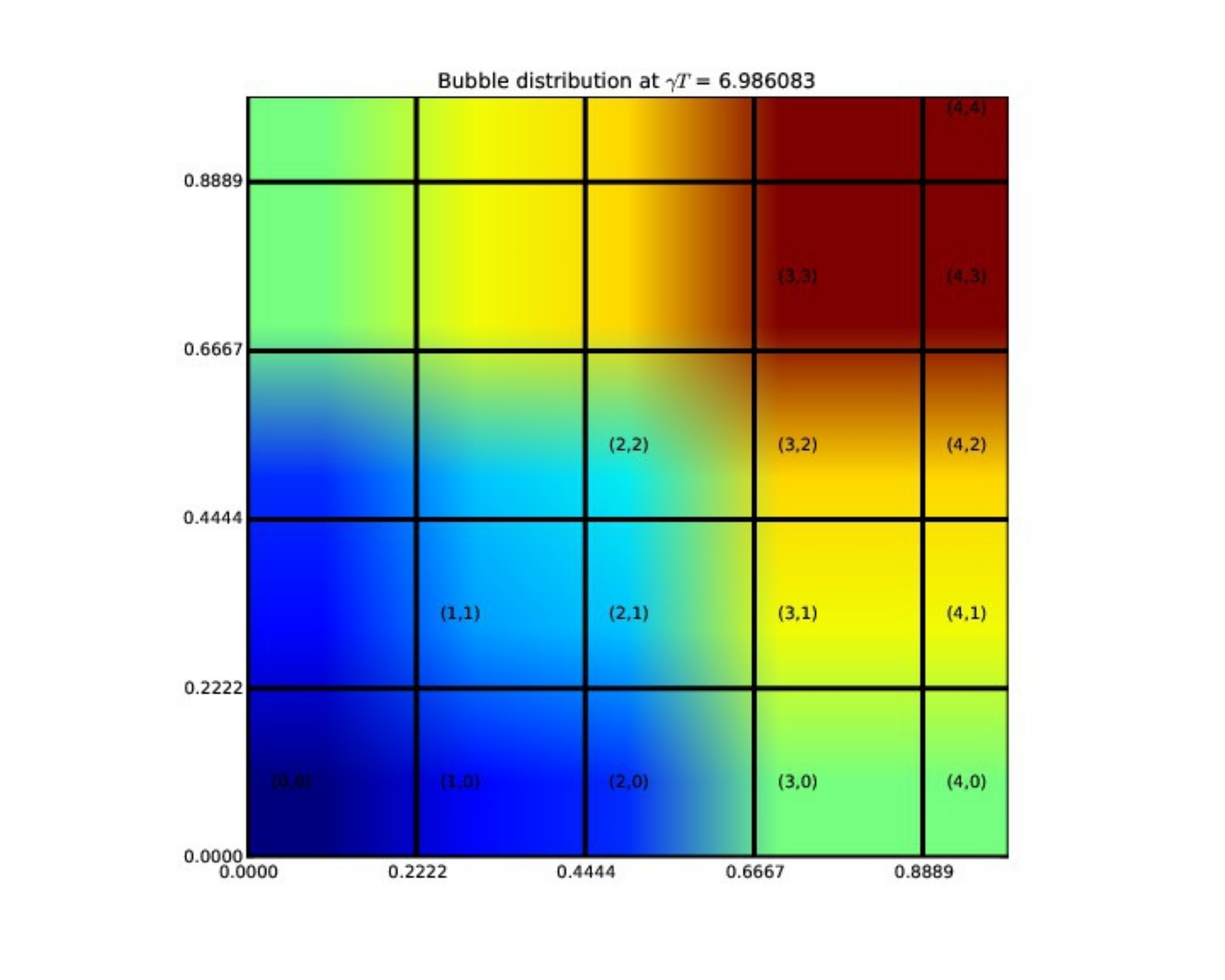}
}
\subfloat[Late time]{
  \includegraphics[width=0.66\columnwidth]{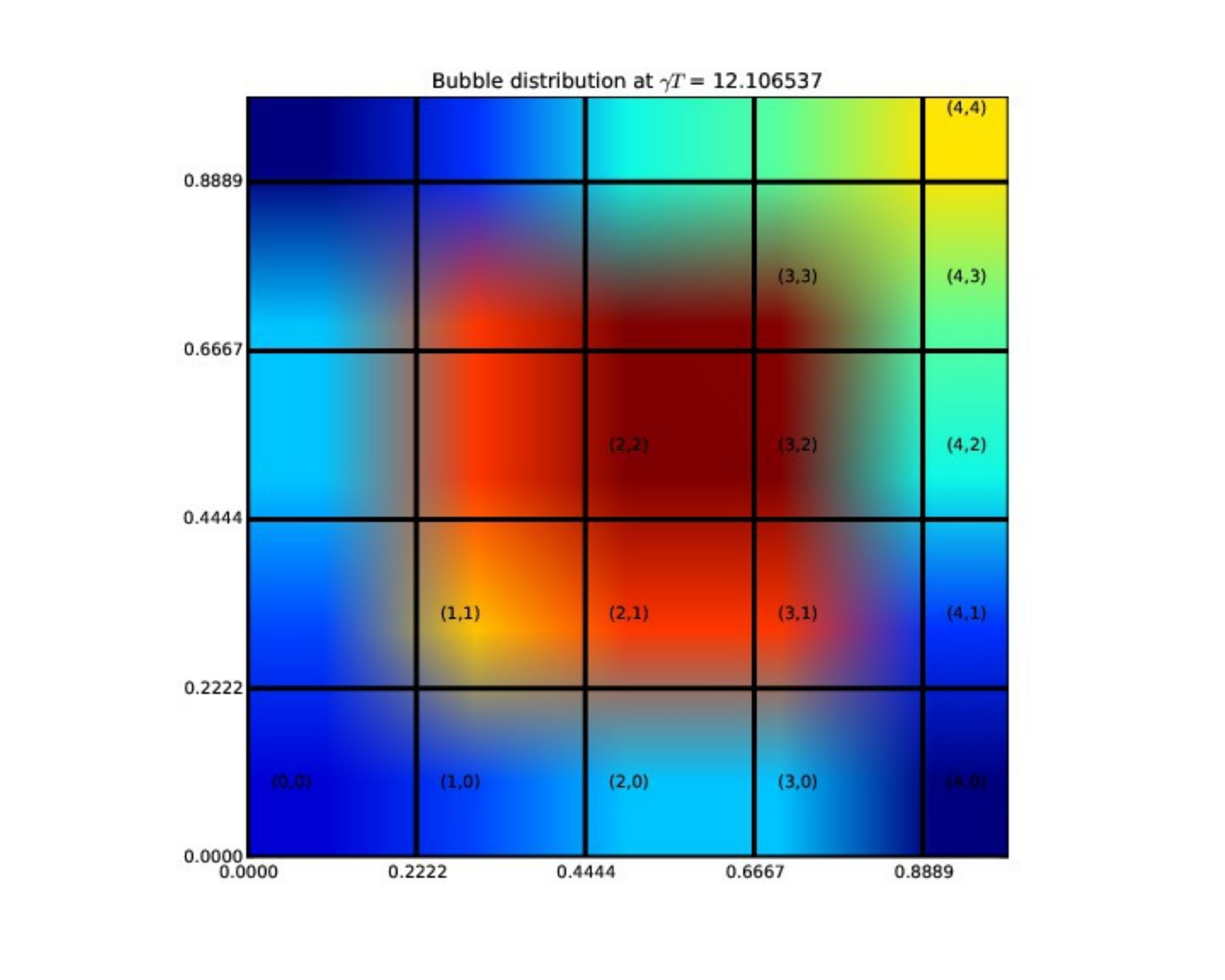}
}
\caption{ \flabel{bubble_slice}
Spatial distribution of bubble heights at three times.
The first time corresponds to the minimum in \fref{ratio_wall}, which is during the transition from linear to non-linear growth
The second time is taken from the stagnation phase.
The third time is taken from the end of the simulation when the central bubbles are squeezed by the growing boundary layer.
}
\end{figure*}

The individual bubble heights for the $4.5$ mode case are computed as in \eref{h_exp}, but restricted to the span-wise domain nearest to the bubble tip, which is marked in \fref{ic}.
The bubble Froude numbers are plotted in \fref{froude_wall}, alongside the aggregate and purely periodic values.

There are at least three mechanisms by which the walls divert the flow from its fully-periodic preference.
The first is that the no-slip boundaries that pass along the diagonal of the boundary bubbles and spikes create a boundary layer that viscously damps vertical flow.
The second is that the pressure gradient from the boundary layer pushes the boundary bubbles and spikes towards the interior of the domain.
These two effects have been studied independently in the context of multi-phase bubbles rising near walls~\cite{Takemura2002}, and are characterized as wall drag and lift forces.
Finally, the finite nature of the span-wise lattice of bubbles and spikes, coupled with the vertical symmetry condition that the total bubble volume must equal the total spike volume, breaks one of the 4-fold symmetries of the infinite cubic lattice causing a local aggregation of spikes in the $(0,0)$ corner and of bubbles in the $(4,4)$ corner.
The local aggregation sets up one low-amplitude long-wavelength mode across the diagonal.

These three mechanisms promote and penalize the growth of different bubbles in the finite lattice, allowing us to infer the relative magnitudes of the effects based on the performance of the bubbles compared to their periodic counterparts.
The wall drag penalizes the growth of bubbles that contain a boundary: the bubbles in the 4th column.
Because the effect also penalizes the growth of the spikes that contain boundaries, it should encourage the growth of the bubbles adjacent to those spikes: those in the 0th row.
The effect should alternate and diminish towards the interior of the domain.
The wall lift pushes bubbles and spikes at the boundary towards the interior.
This reduces the form drag on the interior bubbles by increasing the pressure on their trailing edges.
When adjacent bubbles actually touch, skin drag is also reduced.
Overall, wall lift promotes the growth of the interior bubbles.
Finally, the long-wavelength mode promotes growth of bubbles in the bubble heavy corner and penalizes growth of bubbles in the spike heavy corner.

The spatial distribution of the bubble heights can be seen in \fref{bubble_slice} and the heights relative to the periodic bubble can be seen in \fref{ratio_wall}.
At moderate times, the $(4,4)$ bubble leads and the $(0,0)$ bubble trails, indicating that the long-wavelength mode due to symmetry breaking is the dominant effect.
Additionally, all of the bubbles under-perform their periodic counterpart, indicating that the wall drag has damped the overall flow but with less spatial dependence than the long-wavelength mode.
At late times, the central $(2,2)$, $(2,1)$  bubbles are accelerated while the edge bubbles break down, indicating the growing importance of the wall lift effect.
The wall lift ultimately leads to bubble collisions that destroy the bubble lattice, enhance mixing, and break down the flow.

\subsubsection{Late-time behavior}

\begin{figure*}
\subfloat[Velocity]{
  \includegraphics[width=\columnwidth]{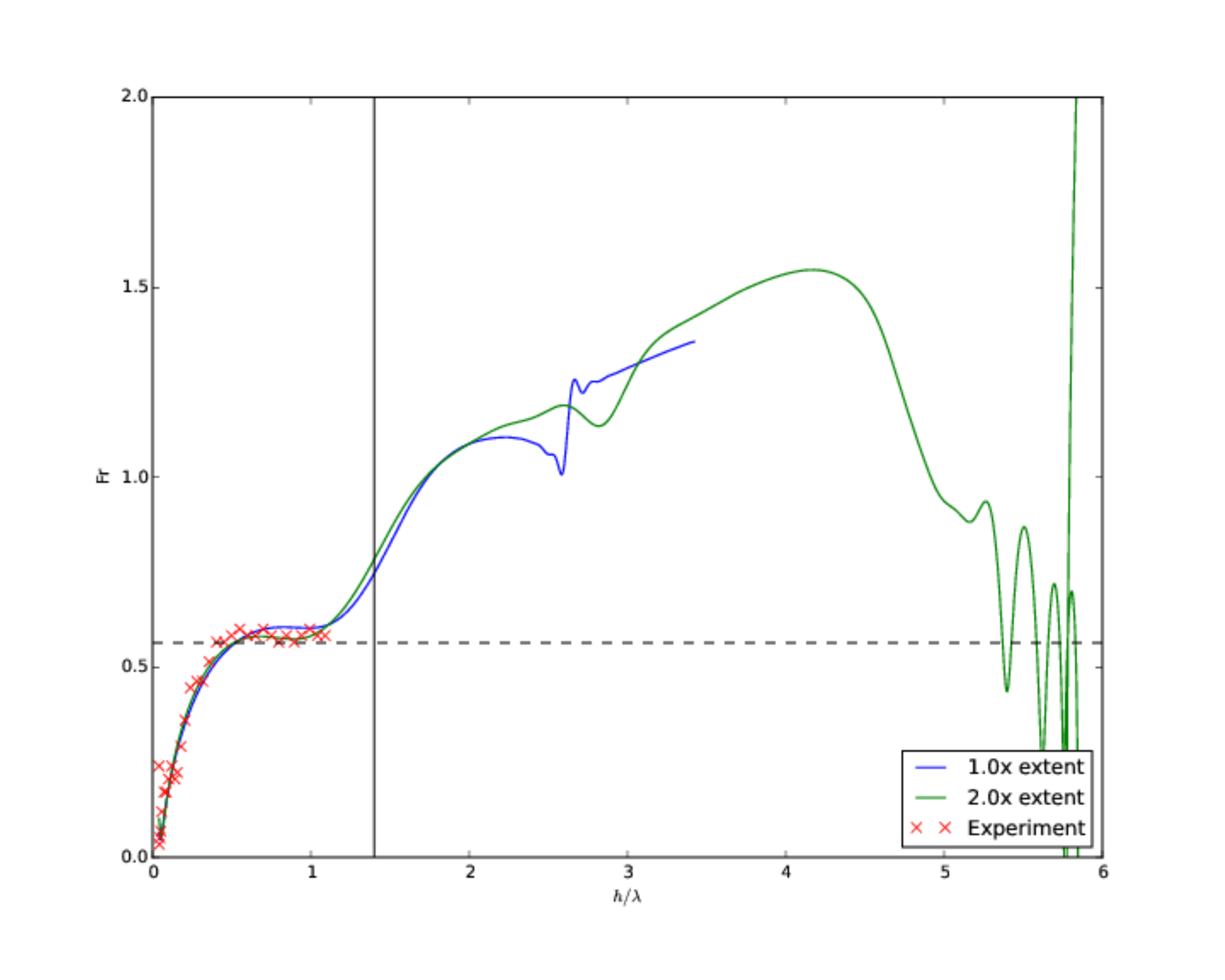}
}
\subfloat[Height]{
  \includegraphics[width=\columnwidth]{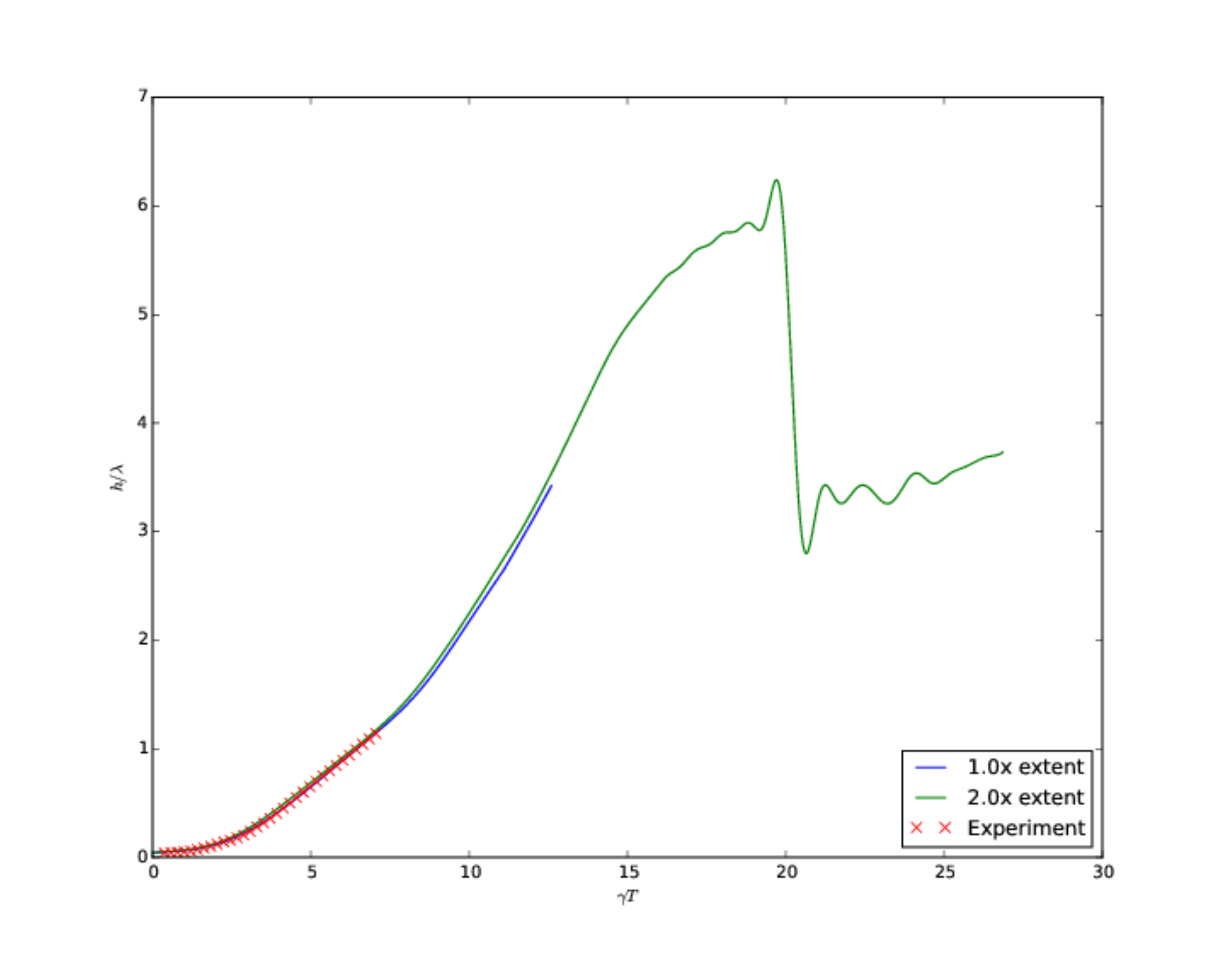}
}
\caption{ \flabel{long_dynamics}
Bubble velocity and bubble height vs time, non-dimensionalized by the wavelength and linear growth rate, for 4.5 mode simulations and experiment.
Lines are from simulation output, one case with the same vertical extent as the simulation and in the other with twice that vertical extent.
Points are from experiment via direct measurement of the bubble velocity and bubble height.
The dotted horizontal line is positioned at Goncharov's theoretical value of $\pi^{-1/2}$~\cite{Goncharov2002}.
The solid vertical line marks the greatest bubble height reached in any of the experiments by Wilkinson and Jacobs~\cite{Wilkinson2007}.
}
\end{figure*}

\begin{figure*}
\subfloat[Full trajectory]{
  \includegraphics[width=0.66\columnwidth]{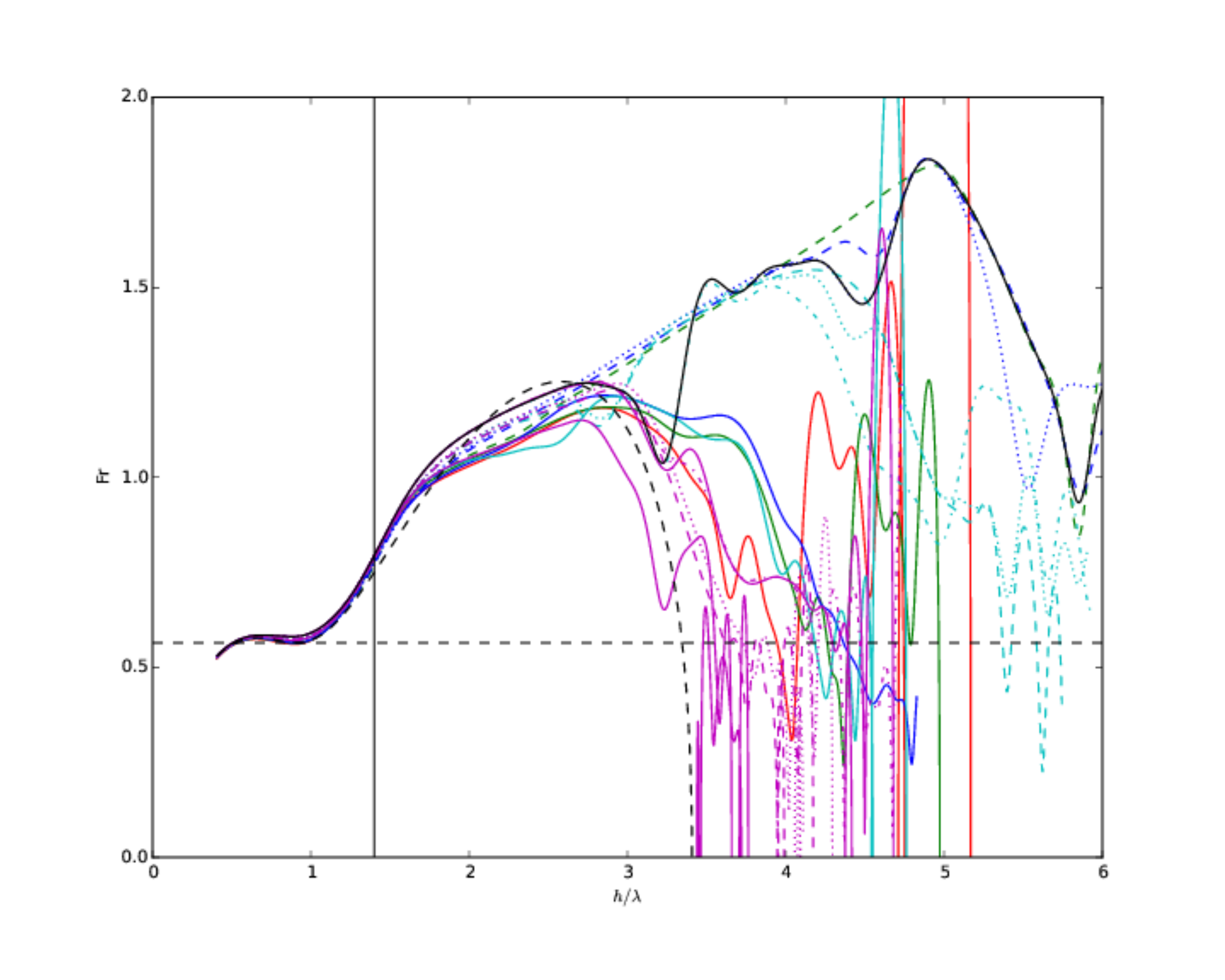}
}
\subfloat[Decay at walls]{
  \includegraphics[width=0.66\columnwidth]{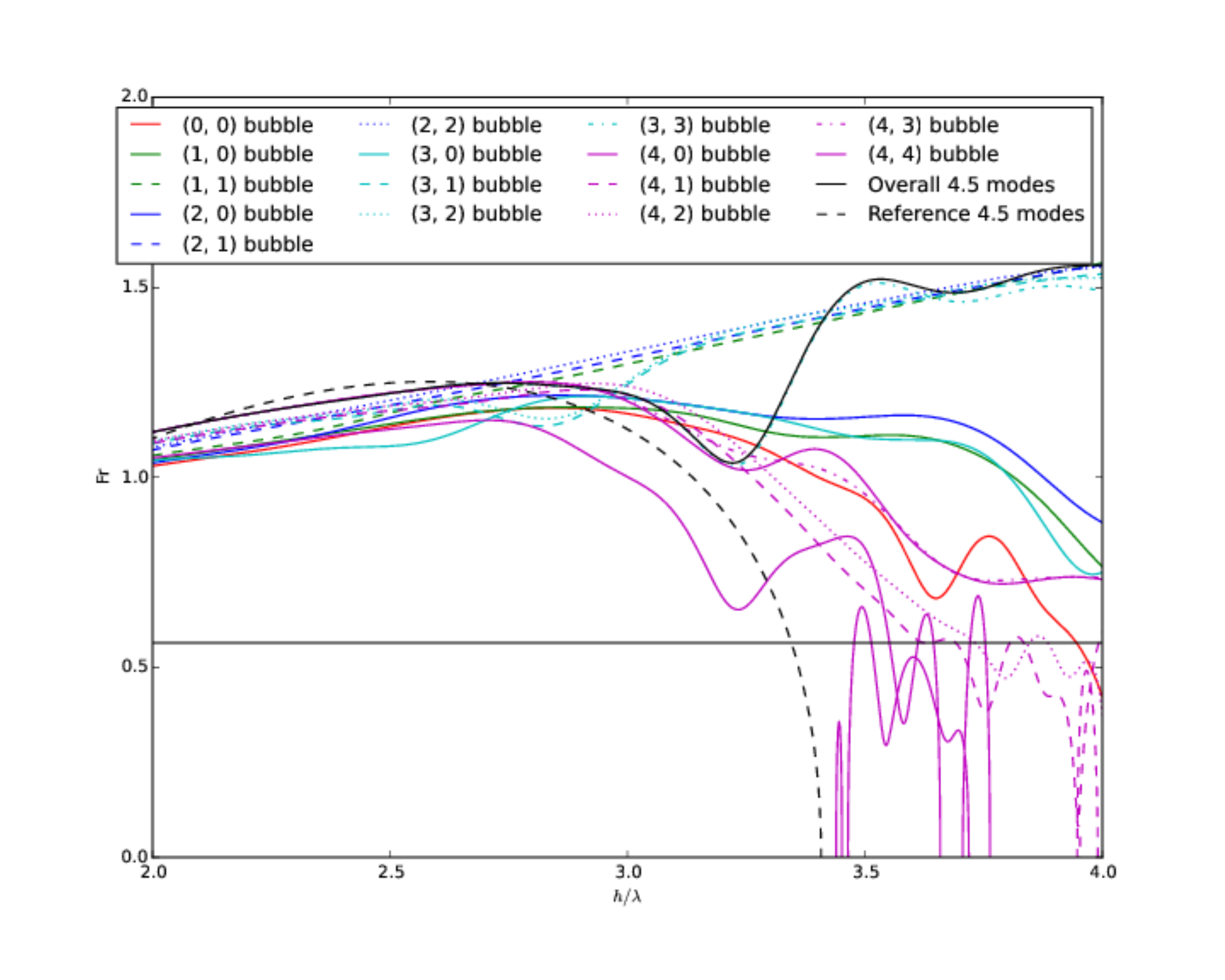}
}
\subfloat[Decay in center]{
  \includegraphics[width=0.66\columnwidth]{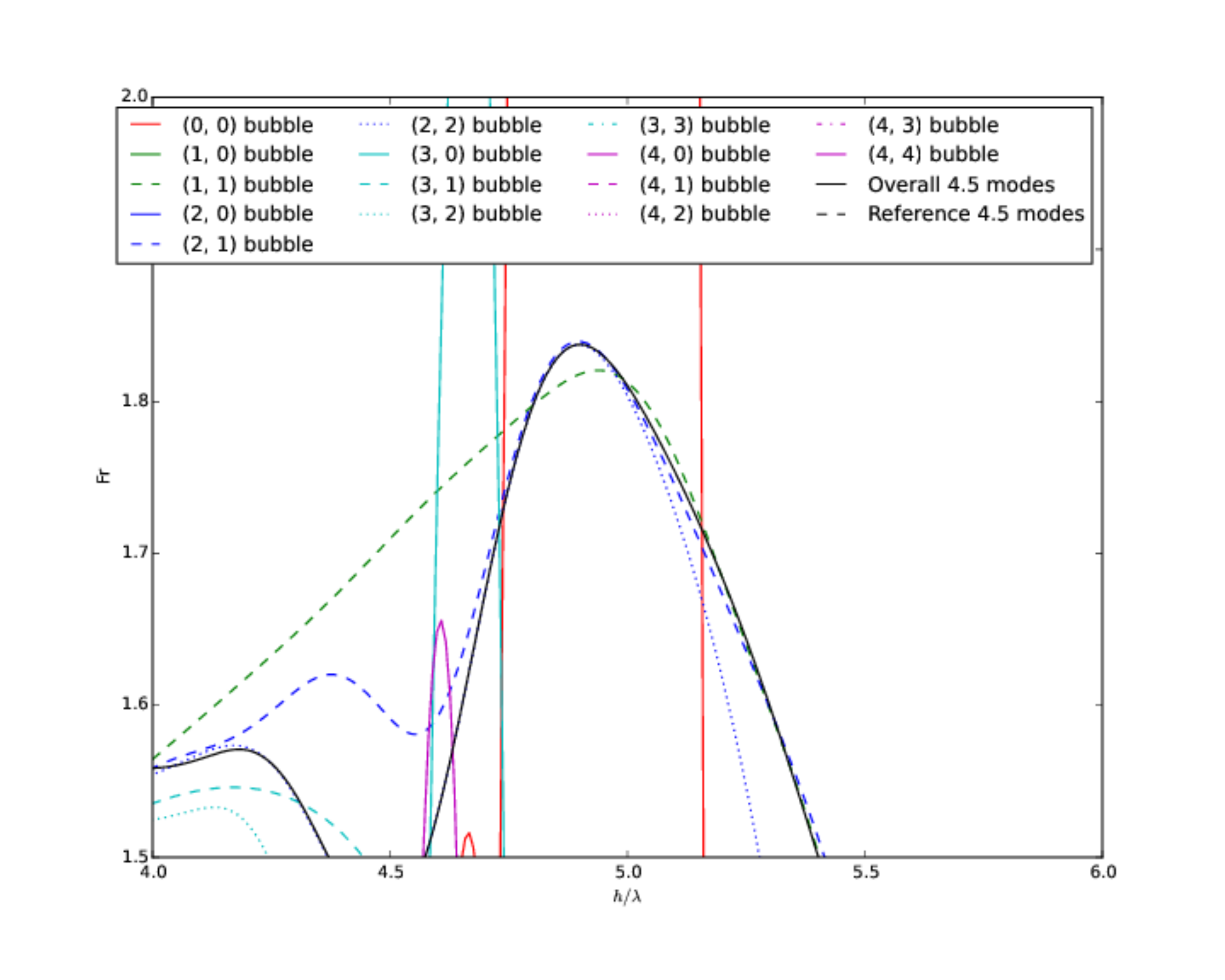}
}
\caption{ \flabel{long_wall_dynamics}
Froude number as a function of height, non-dimensionalized by the wavelength, by bubble in the 4.5 mode simulation with extended vertical exten.
Solid line is from the height defined as the maximum taken over the entire span-wise domain.
Dotted line is the periodic reference calculation.
The dotted horizontal line is positioned at Goncharov's theoretical value of $\pi^{-1/2}$~\cite{Goncharov2002}.
The solid vertical line marks the greatest bubble height reached in any of the experiments by Wilkinson and Jacobs~\cite{Wilkinson2007}.
}
\end{figure*}

The 4.5 mode simulation was repeated with twice the vertical extent and simulation time.
Additionally, the Schmidt number was increased from 1 to 3.5 to reduce late-time mixing not present in high Schmidt number experiments.
\fref{long_dynamics} compares the short unit-Schmidt and long moderate-Schmidt trajectories, which are widely in agreement.
The reduction in bubble acceleration around aspect ratio $h/\lambda = 2$ is present in both the short and the long simulations, so it is unlikely to be due to the vertical domain boundaries.
It is not, however, present in the periodic calculation, so it could be a wall effect.

\fref{long_wall_dynamics} mimics \fref{froude_wall} but for the late-time case.
The periodic reference trajectory, calculated with the original domain size, rapidly decays after reaching a maximum around aspect ratio $h/\lambda = 2.5$ due to interactions with the top of the domain.
Its maximum Froude number is around 1.2, consistent with previous calculations.
The central bubbles in the extended late-time run continue to experience constant acceleration past aspect ratio 3 and Froude number 1.2, with the $(1,1)$ bubble continuing to aspect ratio 5 and Froude number 1.8.

The decay of the velocity of the periodic reference bubble around $h/\lambda = 2.5$ suggests the late-time simulations would interact with the top boundary around $h/\lambda = 5$.
In fact, this is exactly when the $(1,1)$ and $(2,2)$ bubbles begin to decay.
However, the bubbles closer to the boundaries break down much earlier.
The $(4,4)$ bubble, for example, reaching maximum Froude number around $h/\lambda = 2.75$.
We can infer that the wall lift that drives the boundary bubbles into the interior bubbles destroys the periodic ordering.
It is not clear if the decay of the interior bubbles at $h/\lambda = 5$ is due to the top boundary or the wall lift destroying the periodic ordering.

\subsubsection{Secondary flow}

\begin{figure*}
\subfloat[2.5 modes]{
  \includegraphics[width=0.66\columnwidth]{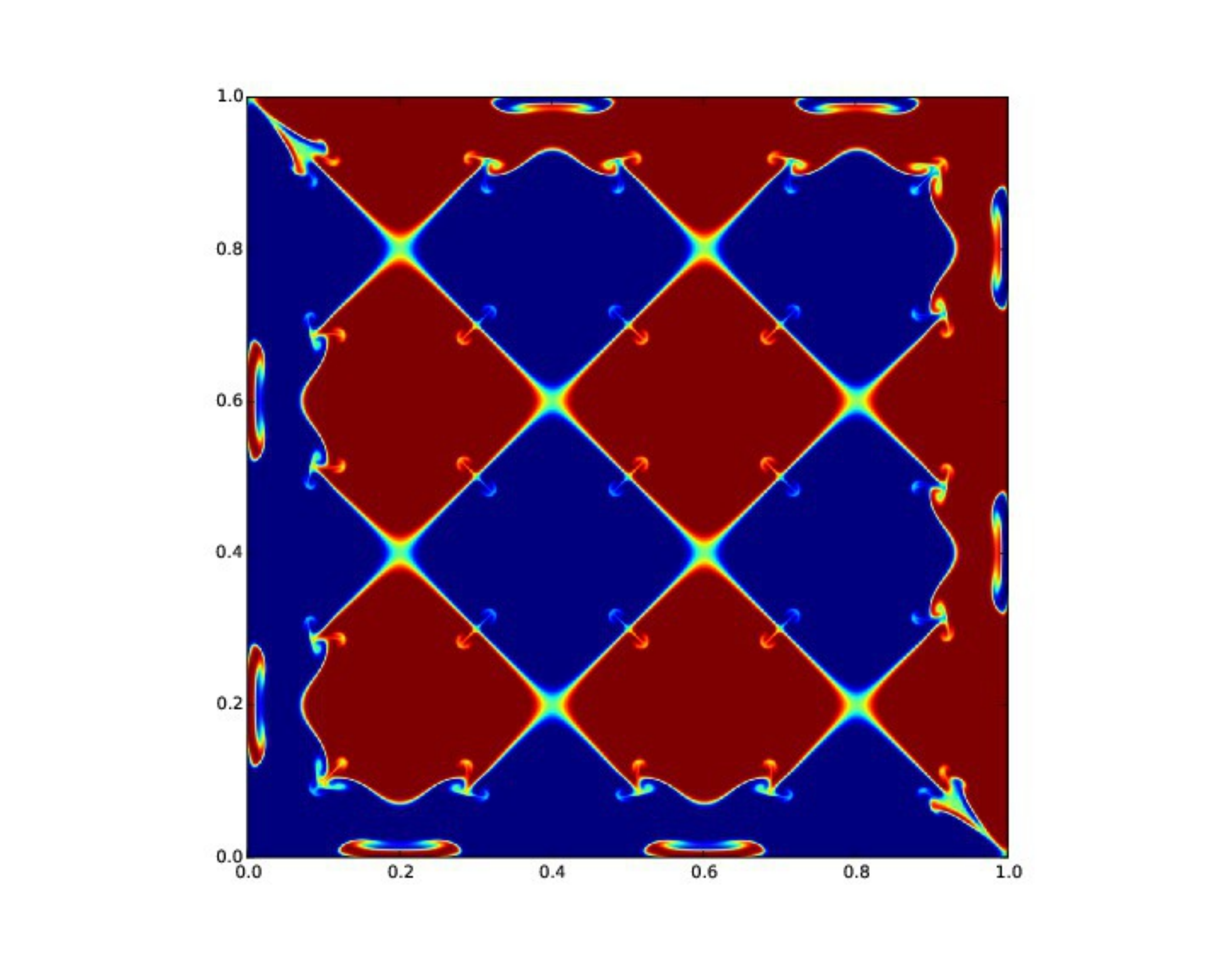}
}
\subfloat[3.5 modes]{
  \includegraphics[width=0.66\columnwidth]{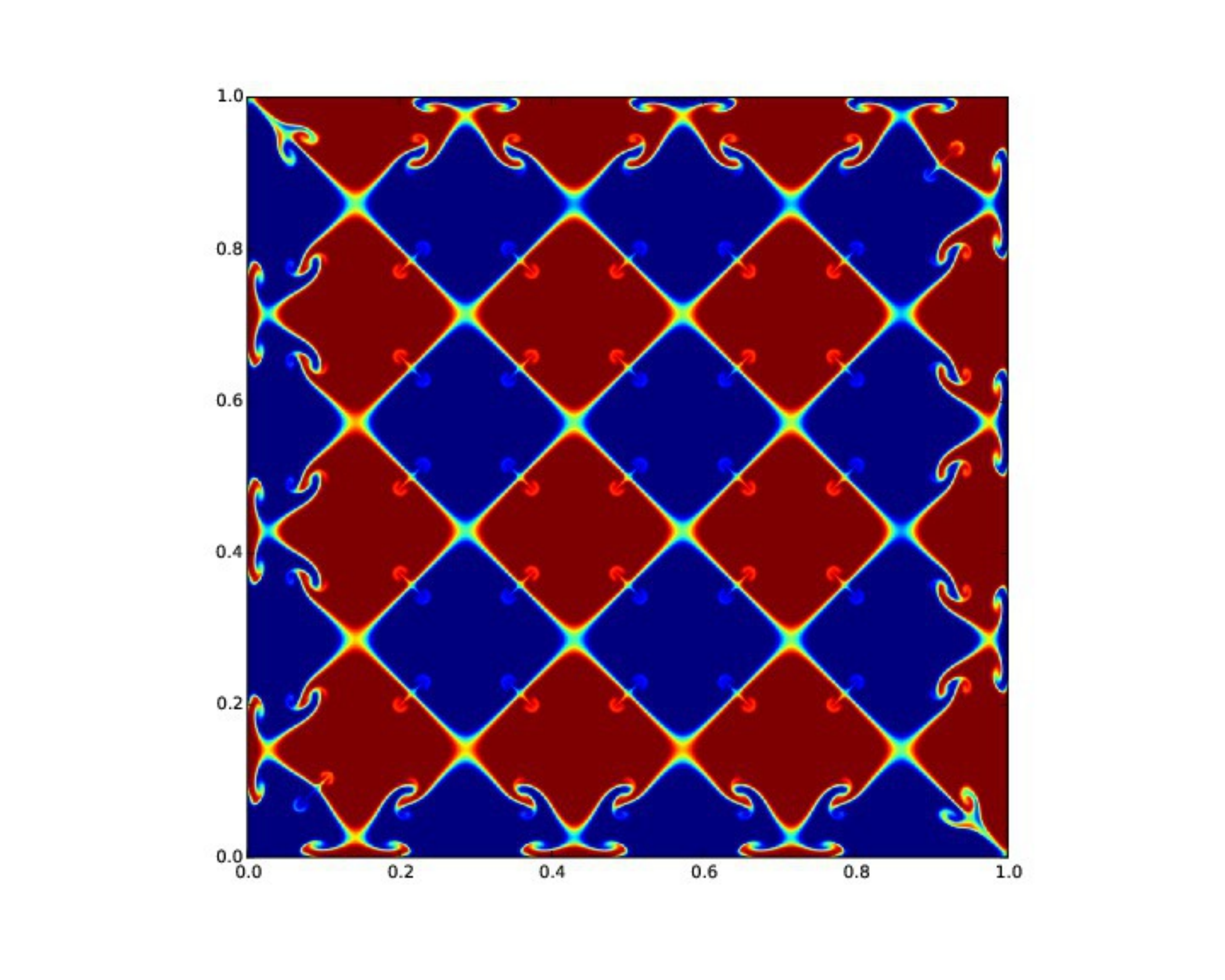}
}
\subfloat[4.5 modes]{
  \includegraphics[width=0.66\columnwidth]{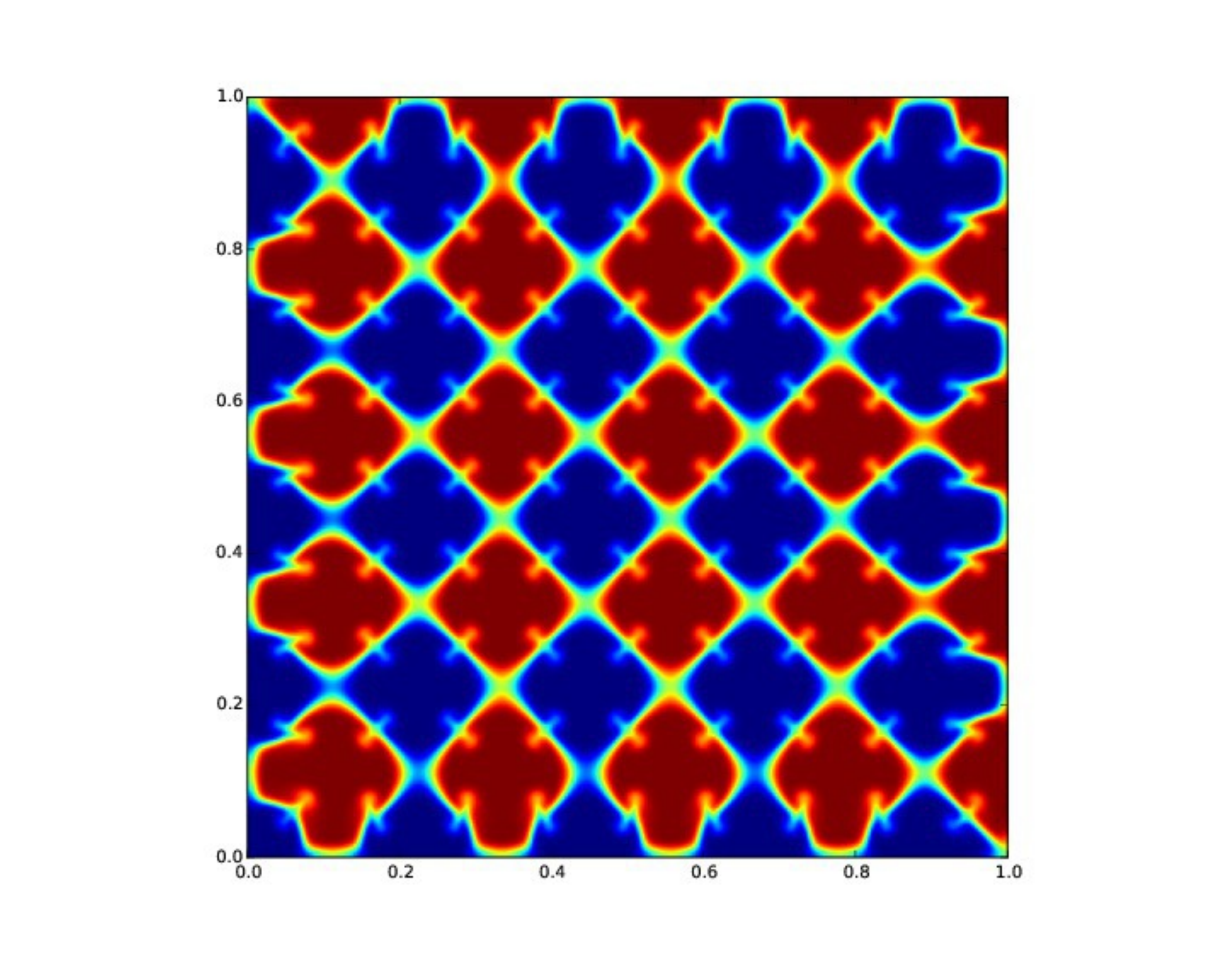}
}
\caption{ \flabel{phixy}
Scalar field in the horizontal mid-plane for 2.5, 3.5, and 4.5 mode simulations.
}
\end{figure*}

\begin{figure*}
\subfloat[2.5 modes]{
  \includegraphics[width=0.66\columnwidth]{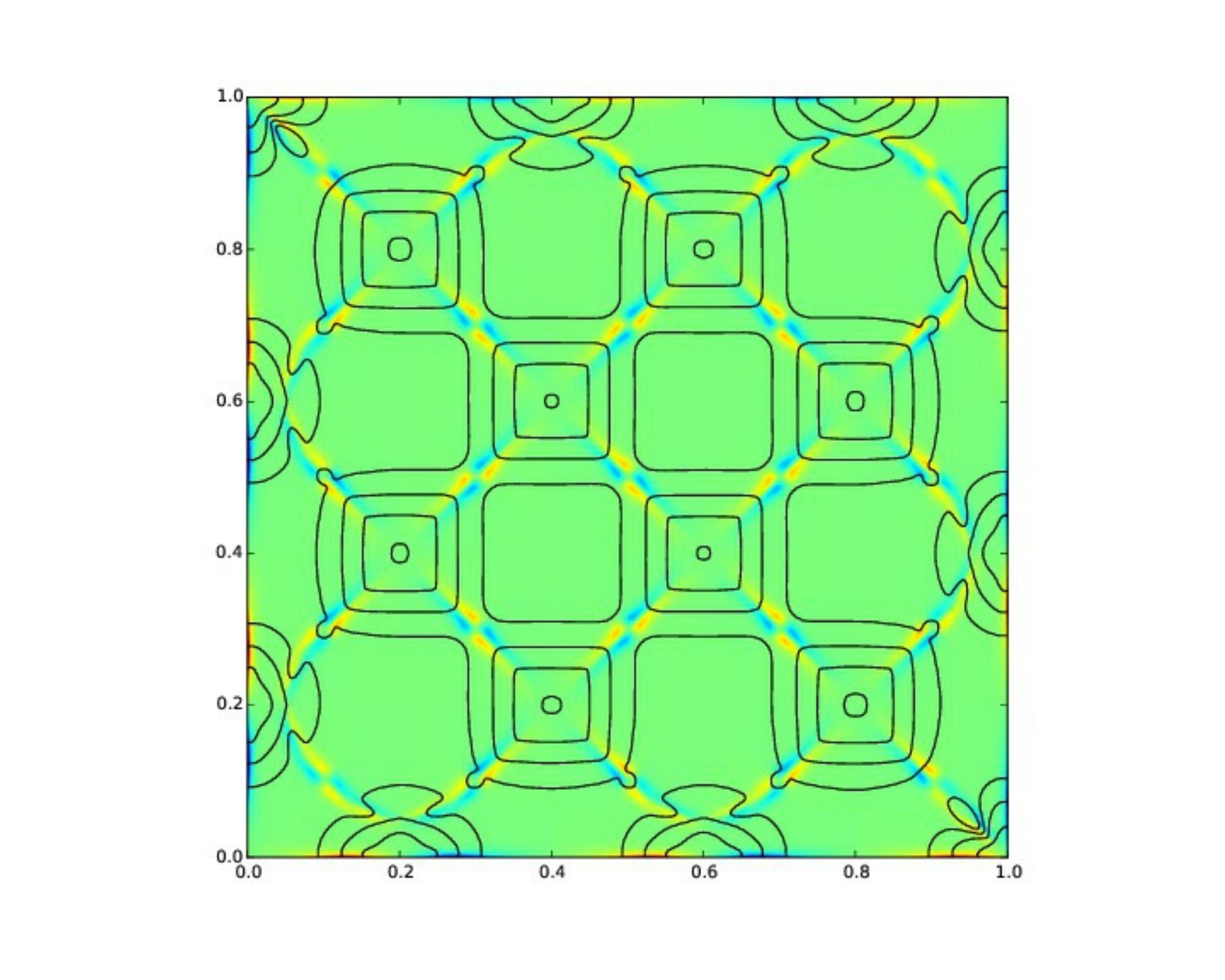}
}
\subfloat[3.5 modes]{
  \includegraphics[width=0.66\columnwidth]{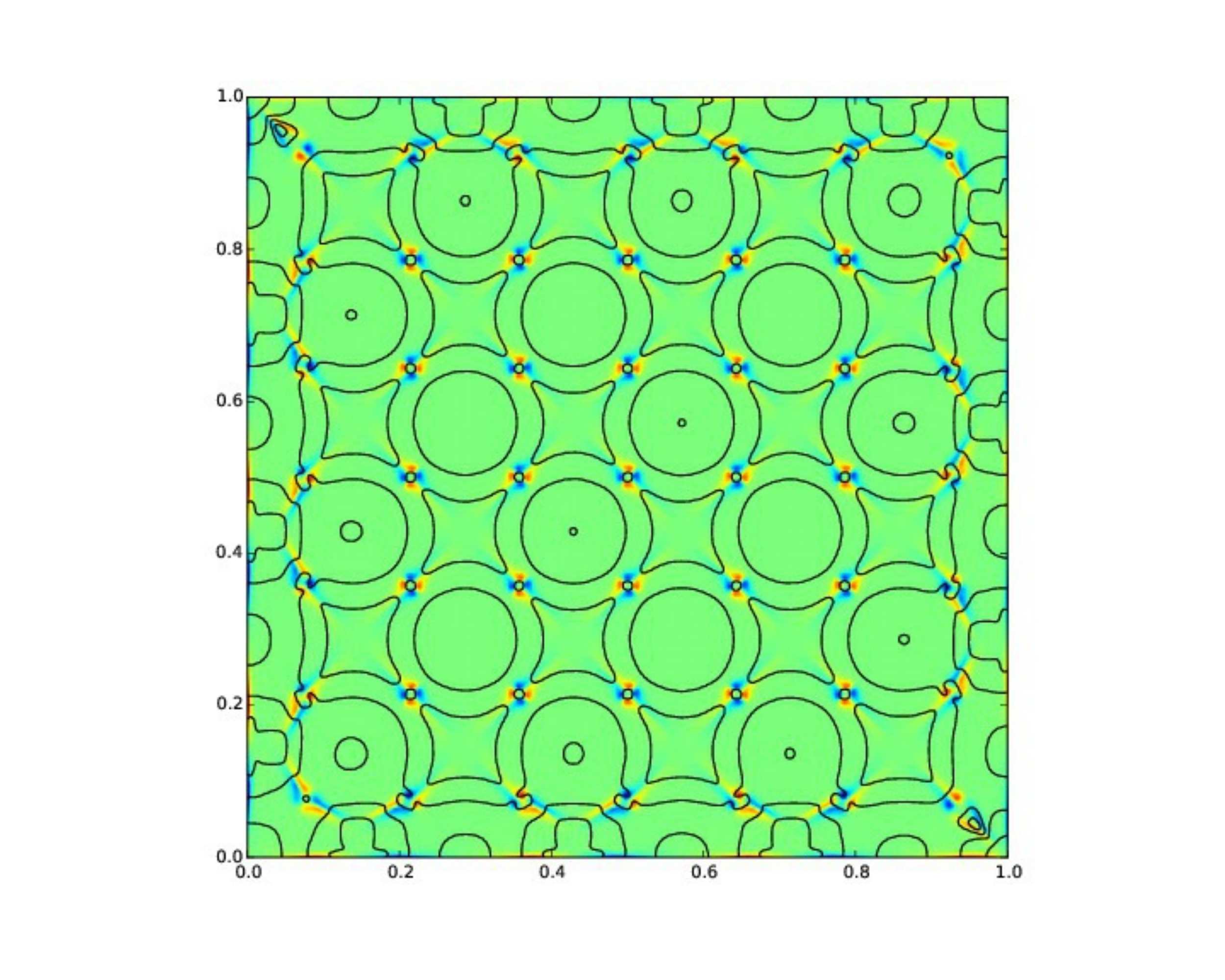}
}
\subfloat[4.5 modes]{
  \includegraphics[width=0.66\columnwidth]{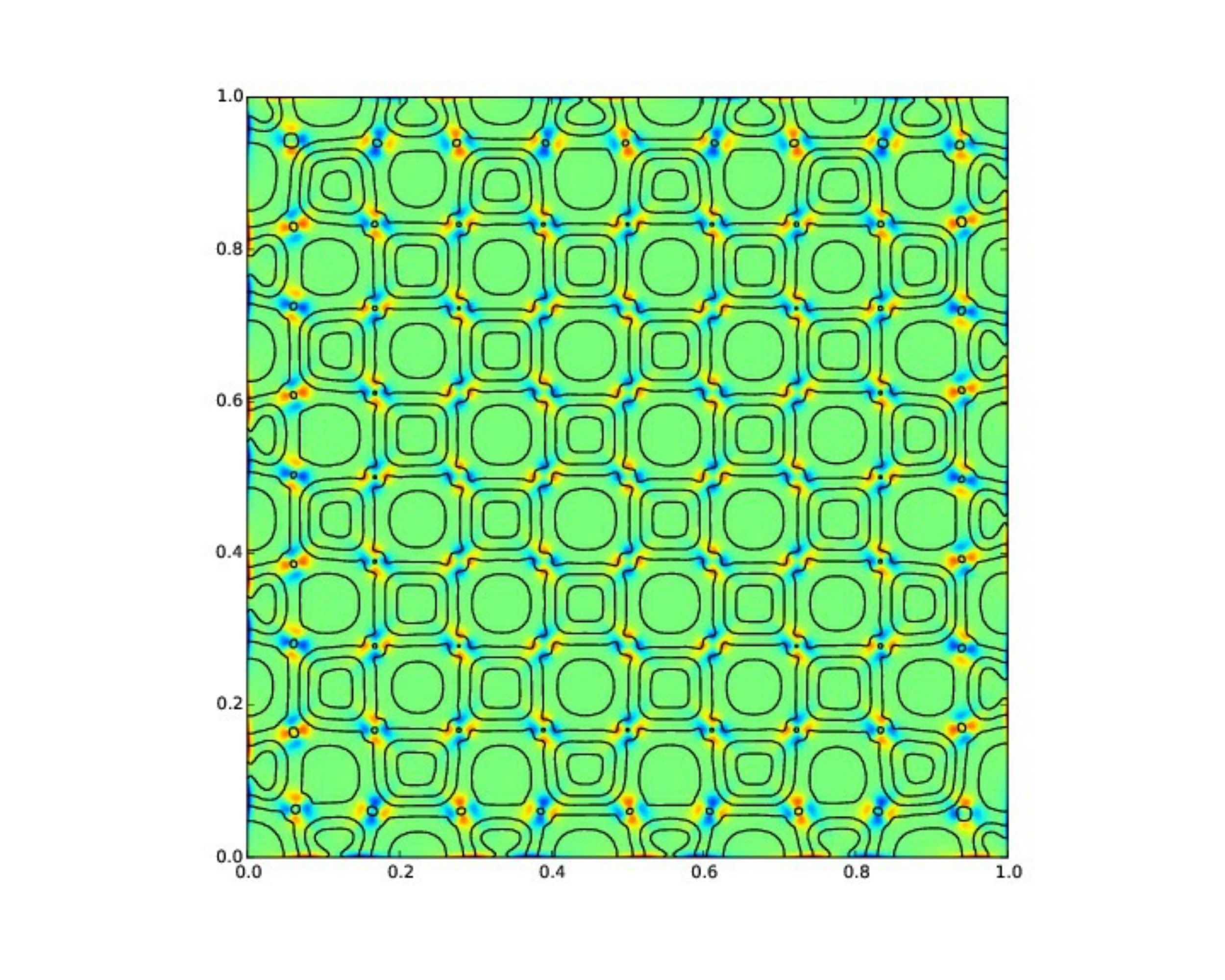}
}
\caption{ \flabel{secondary}
Secondary flow in the horizontal mid-plane.
Background color is the vertical component of the vorticity.
Contours are lines of constant pressure.
}
\end{figure*}

\begin{figure*}
\subfloat[2.5 modes]{
  \includegraphics[width=0.66\columnwidth]{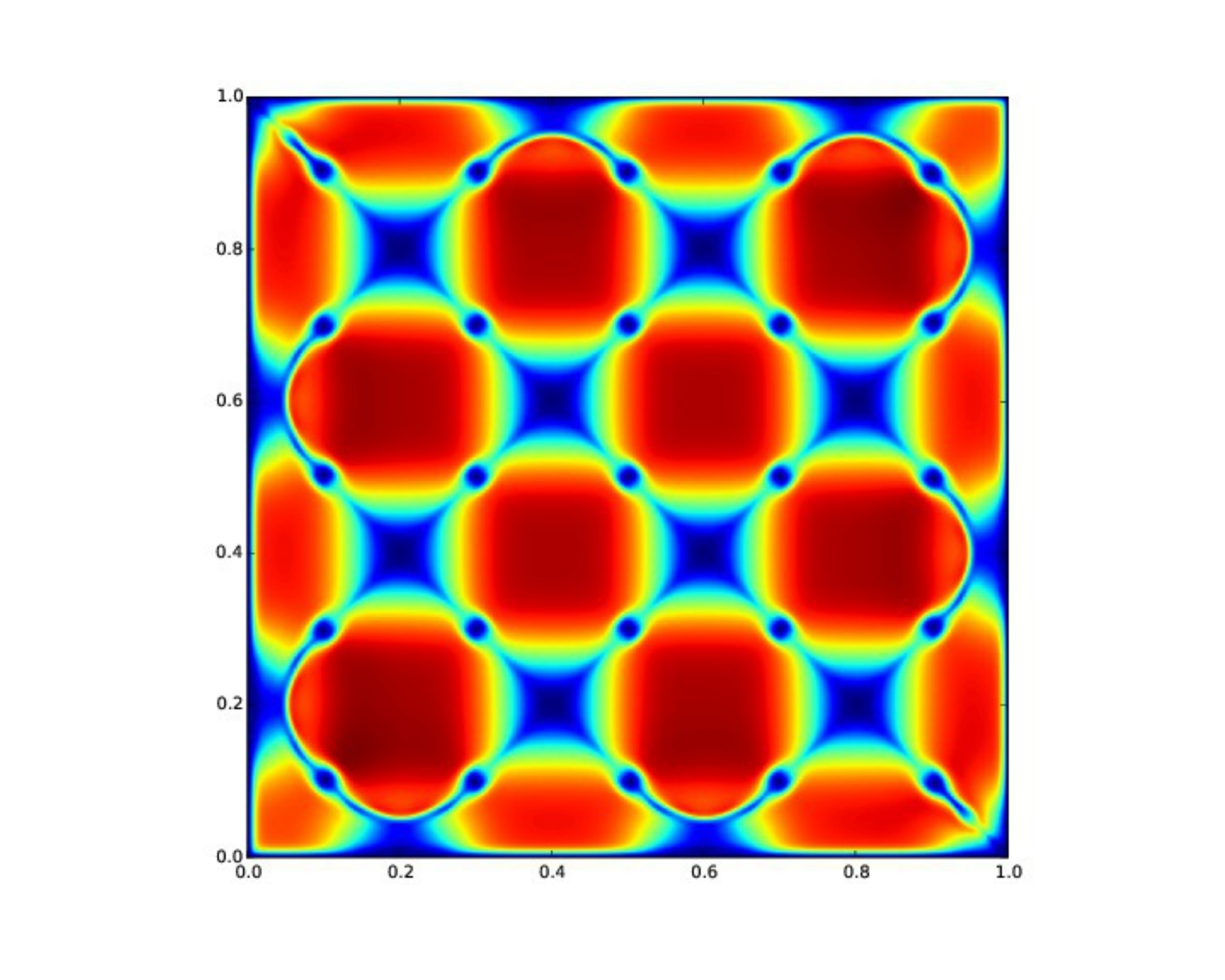}
}
\subfloat[3.5 modes]{
  \includegraphics[width=0.66\columnwidth]{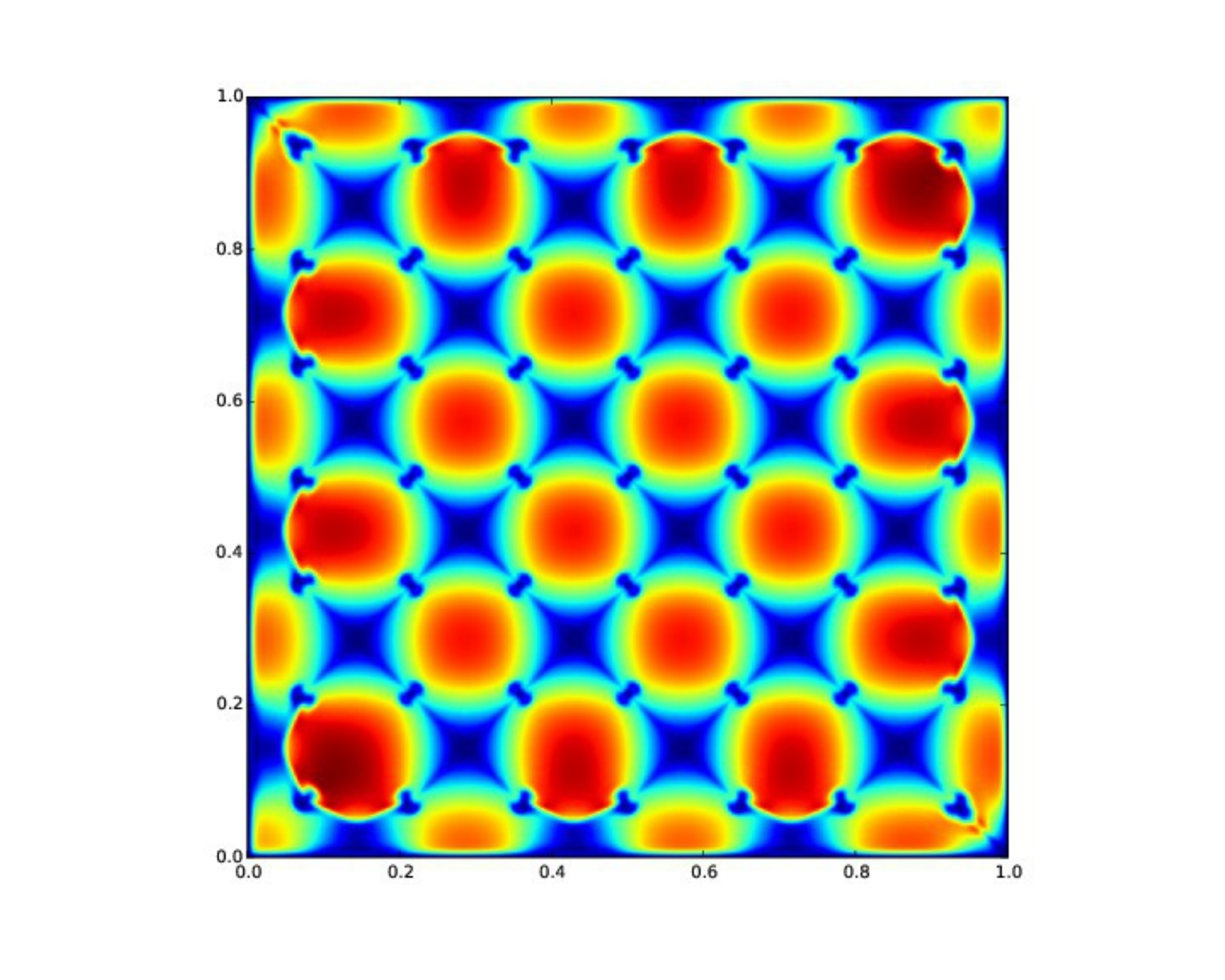}
}
\subfloat[4.5 modes]{
  \includegraphics[width=0.66\columnwidth]{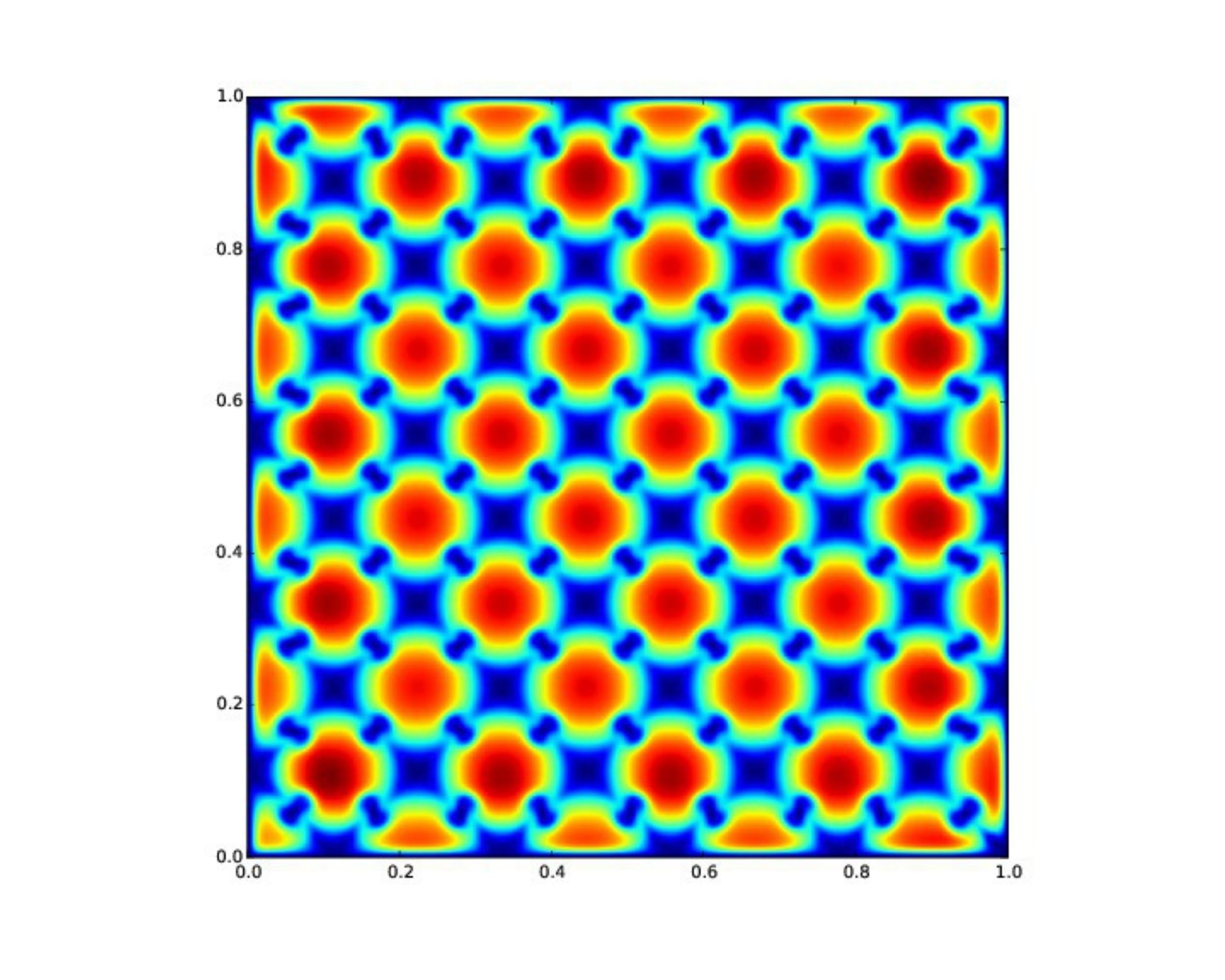}
}
\caption{ \flabel{dynamic}
Dynamic pressure in the horizontal mid-plane.
}
\end{figure*}

In addition to the bubble height and vertical slices in the diagonal plane, we observe the horizontal mid-plane.
The span-wise scalar field, $\phi(x,y,0)$, exhibits plume structures that penetrate the bubble faces, as seen in \fref{phixy}.
The cause of these plumes is advection by secondary flows, as seen in \fref{secondary}.
Overlaying the pressure with the span-wise velocity reveals it to be secondary flow of the first kind: secondary flow due to span-wise pressure gradients.
The span-wise pressure gradient comes from the dynamic pressure of the rising and falling bubbles and spikes, in contrast to the stationary points at their interfaces.

The secondary flow advects mixed fluid from the interface into the centers of the bubbles and spikes.
Enhanced mixing reduces the effective Atwood number of the bubbles and spikes, but the magnitude of this effect is not clear.
As a secondary flow of the first kind, this mixing mode is present even at low Reynolds numbers.

\makeatletter{}\section{Conclusions} \slabel{concs}

The simulations described here reproduce the growth rate, stagnation velocity, and re-acceleration of the low-Atwood single mode Rayleigh Taylor instability for three experimental runs by Wilkinson and Jacobs.
These reproductions inspire confidence not only in the NekBox code, but also in the Boussinesq approximation for $A = 0.15$ and the low-Schmidt approximation.

In wall-bounded flows, the bubbles and spikes nearest to the no-slip boundaries experience lift and drag forces that slow their non-linear growth and push them towards their inner neighbors.
There is an additional effect due to the finite domain breaking one of the 4-fold symmetries from the purely periodic problem.
In the wall-bounded initial condition, one corner of the domain has an excess of bubbles while the opposite has an excess of spikes.
This sets up a long-wavelength mode across the diagonal that encourages bubble growth in one corner and discourages it in the other.

Ultimately, the bubble-bubble and spike-spike collisions may destroy the single-mode ordering of the flow at aspect ratio 5, but the onset of velocity decay may alternatively be due to the upper boundary.
If the decay is due to collisions, it would limit the use of wall-bounded flows as proxies for periodic flows to moderate aspect ratios.
The ability of wall-bounded flows to approximate periodic ones at high aspect ratio warrants further study.

In addition to causing collisions, the growing boundary layer squeezes the flow in the span-wise direction, accelerating it past its fully-periodic trajectory in the stagnation and re-acceleration phases.

The inner bubbles experience near constant acceleration from aspect ratio 2 to aspect ratio 5, with a maximum Froude number of 1.8.
This contrasts results by Ramaprabhu et al.~\cite{Ramaprabhu2012} that show saturation post-reacceleration at $\text{Fr} \approx 1$.
The saturation could be explained by excess mixing or the finite size of the domain, which extends to $h/\lambda = 6$ in their case and $9$ in ours.
Alternatively, the acceleration could be artificially sustained by the wall lift force pushing the boundary bubbles into the interior ones.

Single-mode Rayleigh-Taylor flows develop span-wise pressure gradients with local minima in bubble and spike centers and local maxima in bubble and spike corners.
The pressure drives secondary flows of the first kind in the form of vortex quads centered on bubble-spike interface centers.
These span-wise flows mix the fluid across otherwise laminar interfaces, perturbing the scalar profiles.

\begin{acknowledgements}

M. H. is grateful for useful conversations with Jeffrey Jacobs, Robert Roser, Aleksandr Obabko, Elia Merzari, and Oana Marin.

M. H. acknowledges support from the Department of Energy Computational Science graduate fellowship.
This research used resources of the Argonne Leadership Computing Facility, which is a DOE Office of Science User Facility supported under Contract DE-AC02-06CH11357.

\end{acknowledgements}

\bibliography{library}

\end{document}